\documentclass[manuscript,screen,review=false]{acmart}
\usepackage{pifont}
 \usepackage{todonotes}
\usepackage{mathtools}
\usepackage{physics}
\usepackage{multirow}
\usepackage{wrapfig}
\usepackage{listings}
\definecolor{mygreen}{RGB}{28,172,0}
\DeclareCaptionFormat{listing}
  {\colorbox{gray}
     {\parbox{\dimexpr\textwidth-2\fboxsep}{\centering #1#2#3}}}

\AtBeginDocument{%
  \providecommand\BibTeX{{%
    \normalfont B\kern-0.5em{\scshape i\kern-0.25em b}\kern-0.8em\TeX}}}

\setcopyright{acmcopyright}
\copyrightyear{2022}
\acmYear{2022}
\acmDOI{}

\let\oldtodo\todo
\renewcommand{\todo}[1]{\oldtodo[inline]{{\bf TODO:} #1}}

\newcommand{\donotshow}[1]{}
\newcommand{\fig}{Fig. }
\def\eg{{\em e.g.}, }
\def\ie{{\em i.e.}, }

\newcommand{\blue}[1]{{\color{blue}#1}}

\graphicspath{{FIGS/} {FOTOS/}}

\begin{document}

\title{A CMOS-compatible Ising Machine with Bistable Nodes}

\author{Richard Afoakwa}
\email{richard.afoakwa@rochester.edu}
\affiliation{%
  \institution{University of Rochester}
  \city{Rochester}
  \state{New York}
  \country{USA}
}

\author{Yiqiao Zhang}
\email{yiqiao.zhang@rochester.edu}
\affiliation{%
  \institution{University of Rochester}
  \city{Rochester}
  \state{New York}
  \country{USA}
}

\author{Uday Kumar Reddy Vengalam}
\email{uvengala@rochester.edu}
\affiliation{%
  \institution{University of Rochester}
  \city{Rochester}
  \state{New York}
  \country{USA}
}

\author{Zeljko Ignjatovic}
\email{zeljko.ignjatovic@rochester.edu}
\affiliation{%
  \institution{University of Rochester}
  \city{Rochester}
  \state{New York}
  \country{USA}
}

\author{Michael Huang}
\email{michael.huang@rochester.edu}
\affiliation{%
  \institution{University of Rochester}
  \city{Rochester}
  \state{New York}
  \country{USA}
}


\begin{abstract}\label{sec:abstract}

Physical Ising machines rely on nature to guide a dynamical system towards 
an optimal state which can be read out as a heuristical solution to a
combinatorial optimization problem. Such designs that use nature as a
computing mechanism can lead to higher performance and/or lower operation
costs and hence have attracted research and prototyping efforts from industry 
and academia. Quantum annealers are a prominent example of such efforts. However,
some physics-centric Ising machines require stringent operating conditions 
that result in significant bulk and energy budget.  Such 
disadvantages may be acceptable if these designs provide some significant
intrinsic advantages at a much larger scale in the future, which remains
to be seen. But for now, integrated electronic designs of Ising machines
allow more immediate applications. We propose one such design that uses 
bistable nodes, coupled with programmable and variable strengths. The 
design is fully CMOS compatible for chip-scale applications and demonstrates 
competitive solution quality and significantly superior execution time and 
energy. 

\end{abstract}



\maketitle

\section{Introduction}

The power of computing machinery has improved by orders of magnitude over the
past decades. At the same time, the need for computation has been spurred
by the improvement and continues to require better mechanisms to solve a
wide array of modern problems. For a long time, the industry focused on
improving general-purpose systems. In recent years, special-purpose designs
have been increasingly adopted for their efficacy in certain type of tasks
such as encryption and network operations~\cite{5434075, erbagci_isscc2015, 
song_isscc2018, satpathy_isscc2016}. More recently, machine learning tasks 
have become a new focus and many specialized architectures are proposed to 
accelerate these operations~\cite{dadiannao2014,Jouppi_cm2017}. Much of 
this work is to construct a more efficient architecture where the control 
overhead as well as the cost of operation becomes much lower than traditional 
designs.

In a related but different track of work, researchers are trying to map
an entire algorithm to physical processes such that the resulting state
represents an answer to the mapped algorithm. Quantum computers marketed by
D-Wave Systems are prominent examples. Different from circuit model quantum
computers~\cite{google_quantum,pednault2019leveraging},
D-Wave machines perform quantum
annealing~\cite{bunyk_2014_tas}.\footnote{
Recent theoretical works have claimed increasingly strong equivalence between
the two modes of quantum computing~\cite{Yu_2018, Dam2001HowPI}.}
The idea is to map a combinatorial
optimization problem to a system of qubits such that the system's energy
maps to the metric of minimization. Then, when the system is controlled
to settle down to the ground state, the state of qubits can be read out, which corresponds
to the solution of the mapped problem.

It is as yet not definitive whether D-Wave's systems can reach some sort of
quantum speedup. But one thing is clear: machines like these can indeed find
\emph{some} good solutions to an optimization problem, and in a very short
amount of time too. Indeed, a number of alternative designs have emerged
recently all showing good quality solutions for non-trivial sizes (sometimes
discovering better results than the best known answer from all prior attempts)
in milli- or micro-second latencies~\cite{inagaki2016cim, wang2019oim, carmes_comm2020}.
These systems all share the property that a problem can be mapped to the
machine's setup and then the machine's state evolves according to the physics
of the system. This evolution has the effect of optimizing a particular
formula called the Ising model (more on that later). Reading out the state
of such a system at the end of the evolution thus has the effect of obtaining
a solution (usually a very good one) to the problem mapped.

For example, in some systems, the Hamiltonian is closely related to the Ising
formula. Naturally, the system seeks to enter a low-energy state. In other
systems, a Lyapunov function of the system can be shown to be related to the
Ising formula. In general, these systems can be thought of as optimizing an
objective function (in the form of the Ising formula) due to physics.
Hence, they are generally referred to as \emph{Ising machines}. Clearly,
unlike in a von Neumann machine, there is no explicit algorithm to follow.
Instead, nature is effectively carrying out the computation.
Ising machines have been implemented in a variety of ways with very different
(and often complex) physics principles involved. It is unclear (to us at least)
whether any particular form has a fundamental advantage that will manifest in
a very large scale.

Note that these systems can not guarantee reaching the ground state in
practice.\footnote{Theoretical guarantee in some ideal setup may exist. For
instance, adiabatic quantum computing theory says that when the annealing
schedule is \emph{sufficiently} slow and in the absence of noise (zero Kelvin)
the system is guaranteed to stay in the ground state~\cite{messiah_qm}.}
Nonetheless, some systems find a good answer with high speed and a good energy
efficiency, as we shall see later with concrete examples.
In this paper, we propose a novel CMOS-compatible Ising machine which uses circuit
elements' physical properties to achieve nature-based computation. Our proposed 
design is completely different from other efforts of using CMOS circuit
to build machines that \emph{simulate} an annealer, such as chip-based accelerated simulated annealers~\cite{Kirkpatrick671,yamaoka_isscc2015,takemoto_isscc2019}. 
We perform a detailed
analysis of the design and show that it is a compelling design and superior
in many respects to existing Ising machines and accelerators of simulated
annealing. 

\section{Background and Related Work}
\label{sec:background}

We first explain the background of Ising machines and discuss the state of 
the art in implementations.

\subsection{Ising model}\label{sub:ising_model}

The Ising model is used to describe the Hamiltonian of a system of
spins.\footnote{Though commonly called the Ising model, the model itself
existed before Ernst Ising (read ``Easing") solved analytically a
one-dimensional system.}
The model is a general one that describes a system with many nodes (\eg
atoms), each with a spin ($\sigma_i$) which takes one of two
values ($+1$, $-1$). The energy of the system is a function of pair-wise
coupling of the spins ($J_{ij}$) and each spin's
reaction ($h_i$) to some external magnetic
field ($\mu$). The resulting Hamiltonian is as follows:
\begin{equation}
\label{eqn:Ising_w_field}
H = -\sum_{\mathclap{(i<j)}} J_{ij}\sigma_i\sigma_j - \mu \sum_{\mathclap{i}} h_i\sigma_i
\end{equation}
If we ignore the external field, the Hamiltonian simplifies to
\begin{equation}
\label{eqn:Ising}
H=-\sum_{\mathclap{(i<j)}} J_{ij}\sigma_i\sigma_j 
\end{equation}
This simplified version is more useful for the purpose of our discussion. 
Henceforth, when we refer to the Ising model or formula, we mean Eq.~\ref{eqn:Ising}.

A physical system with such a Hamiltonian naturally tends towards low-energy
states and thus serves as a convenient machine to solve a problem with a
formulation equivalent to the Ising formula -- provided we can configure
parameters (\eg $J_{ij}$) to match that of the problem.

\subsection{Optimization problems and mapping issues}\label{sub:optimization}

A group of optimization problems naturally map to an Ising machine. Perhaps
the most straightforward problem to map is (weighted) Max-Cut. Given a 
graph, $G\,=\,(V,E)$, a \emph{cut} is a partition of vertices into two sets 
of, say, $V^{+}$ and $V^{-}$, where $V^{-} = V-V^{+}$.

The Max-Cut problem tries to find a cut such that the combined weight of the 
edges spanning the two sets of vertices is maximum. In other words, the maximum 
cut is
\begin{equation}\label{eqn:max_cut}
    \underset{V^+\in \mathbb{P}(V)}{\operatorname*{arg\,max}}\left(\hskip 8ex
    \sum_{\mathclap{(i,j)\in E;\  i\in V^{+};\ j\in V^{-}}}  W_{ij}\hskip 4ex \right)
    \end{equation}
where $W_{ij}$ is the weight of edge $(i,j)$. (We will refer to the resulting
$\sum W_{ij}$ as the cut \emph{value} in this paper.)

It is easy to see the resemblance between Eq.~\ref{eqn:Ising} and \ref{eqn:max_cut}.
In fact, if we set the coupling weight ($J_{ij}$) to be the negative of
edge weight ($-W_{ij}$) then the Ising formula is simply twice the negative
cut value plus a problem-specific constant ($\sum W_{ij}$) as follows 
(for notational simplicity, for $i\ge j$  we set $W_{ij}$ to 0):

\begin{equation}
\begin{aligned}
H & = -\sum \mathclap{} J_{ij}\sigma_i\sigma_j = \sum_{\mathclap{\sigma_i = -\sigma_j}} W_{ij}\sigma_i\sigma_j + \sum_{\mathclap{\sigma_i = \sigma_j}} W_{ij}\sigma_i\sigma_j \\
  & = -\sum_{\mathclap{\sigma_i = -\sigma_j}} W_{ij} + \sum_{\mathclap{\sigma_i = \sigma_j}} W_{ij} = -2 \sum_{\mathclap{\sigma_i =-\sigma_j}} W_{ij} + \sum W_{ij}
\end{aligned}
\end{equation}

Hence if the machine finds the ground state of the
Hamiltonian, we have the maximum cut. Finding out the maximum cut of an arbitrary
graph is an NP-hard problem. Practical algorithms only try to find a good
answer. Similarly, existing Ising machines (including our design) are all
Ising \emph{sampling} machines that typically provide a good 
sample of a low-energy state, with no guarantee of optimality.

Because of the trivial mapping of the Max-Cut problem to the Ising formula,
designers of Ising machines, often focus on this optimization
problem. However, other optimization problems can also be mapped to an
Ising machine. Indeed, every problem in the original 
NP-complete set~\cite{Karp1972} 
can be expressed by an Ising formulation specifically designed for that 
problem~\cite{lucas2014ising}. Note that Ising formulation may require
more nodes than that of the original formulation and usually requires
additional time to compute coupling coefficients in the Ising formulation
from the original formulation. This transformation is largely straightforward
and the need for it is problem-dependent and thus shared by all Ising machines.

\begin{figure}[htp]
    \centerline{\includegraphics[width=0.8\textwidth]{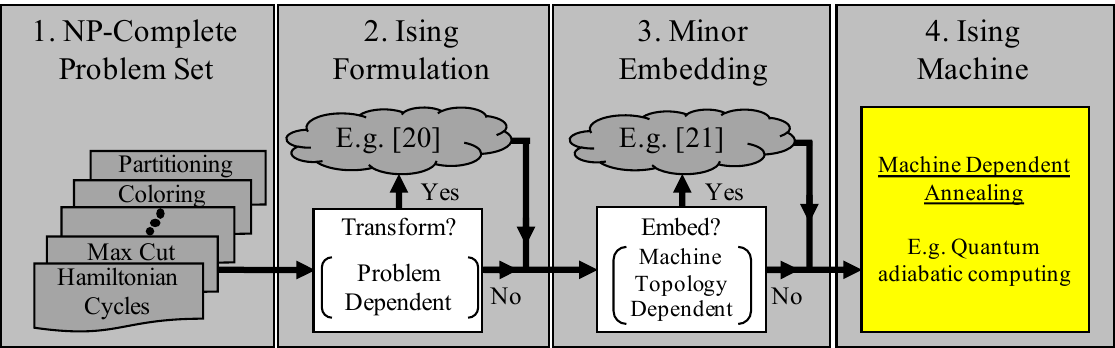}}
    \caption{High-level overview of a generic Ising machine workflow.
    Ising formulation (box-2) is a required property for solving diverse 
    problem sets. Minor-embedding (box-3) is a topology-dependent design 
    factor and \emph{not} required for all Ising machines. The yellow box 
    represents the machine dynamics.
    }
    \label{fig:ising_machine}
\end{figure}

Another transformation, however, may be necessary depending on the machine's
coupling topology. While we will get into the details as we discuss the
machines, it is worth emphasising up front the significant impact of the
issue. If a machine does not have all-to-all connections between physical
nodes, two coupled logical nodes may be mapped to physical nodes (call them
$A$ and $B$) not directly connected. As a result, additional auxiliary
physical nodes are needed, through which nodes $A$ and $B$ are eventually
connected. An alternative description is that if a machine has a limited
connection topology, then the graph of a problem needs to be transformed (\eg
using minor embedding~\cite{cai2014practical}) into a new graph that observes
the limitation imposed by the machine. Consequently, a graph of size $N$
may contain many more nodes (\eg $\frac{N^2}{2}$) after the transformation.
\fig~\ref{fig:ising_machine} illustrates these transformations in the process
of solving a problem on a generic Ising machine.

\subsection{Quantum mechanical and optical Ising  machines}\label{sub:quantum_optical}

There are many natural systems that can be described by the Ising model.
We take two such existing systems
for example. D-Wave's quantum annealers 
use superconducting qubits as the basic building block.
These bits are then coupled together with couplers forming a connection topology
known as the Chimera graph. This is an important architectural constraint
that limits the topology of the problem that can be mapped to the machine.
As we will see later, despite supporting nominally more than 2000 spins,
many of our benchmarks can not be mapped to the machine.
Another disadvantage of the system is the cryogenic operating condition (15mK) needed
for the quantum annealer. This requirement consumes a significant portion of
the 25KW power of the machine~\cite{dwave_stats}.

Coherent Ising machines (CIMs) are another recent example of Ising sampling
machines~\cite{inagaki2016cim, cim_yamamoto, McMahon614, 16bit_cim, 
ncomm_bohm2019}. In a CIM, an optical device called OPO 
(optical parametric oscillator) is used to generate and manipulate the
signal to represent one spin. Unlike in a D-Wave Ising machine, the coupling
between spins in CIM is relatively straightforward in principle. As a result,
CIM implementations have always supported all-to-all coupling. The authors
also emphasized that the 2000-node CIM is therefore far more capable than D-Wave
2000Q which can only map problems of size 64~\cite{Hamerly2019ExperimentalIO}. 

CIM is not without its disadvantages. To support 2000 spins,
kilometers of fibers are needed. Temperature \emph{stability} of the system is
thus an acute engineering challenge. Efforts to scale beyond the currently
achieved size (of about 2000) have not been successful as the system runs into
stability problems. Also worth noting is that the coupling between nodes is --
at least in the current incarnation -- implemented via computation external
to the optical cavity. There is a rather intensive computational demand 
(100s of GFLOPS)~\cite{Hiroki_JPSJ_2019}.
Every pulse's amplitude and phase are detected and its
interaction with all other pulses calculated on an auxiliary computer (FPGA).
The computation is then used to modulate new pulses that are injected back into 
the cavity. Strictly speaking, the current implementation is a nature-simulation 
hybrid Ising machine. Thus, beyond the challenge of constructing the cavity, CIM 
also requires a significant supporting structure that involves fast conversions
between optical and electrical signals. 

These room-sized Ising machines are certainly worthwhile creations
for the sake of science. In particular, investigations are needed to 
see whether the theoretical
underpinning for these machines is relevant in practice.
As we shall see later, both models have significant room for improvements.

\subsection{Electronic oscillator-based Ising machines}\label{sub:oscillator}

A network of coupled oscillators is another physical implementation of an
Ising machine. After
sufficient time, the coupled oscillators will synchronize forming stable relative
phase relationship.\footnote{The observation of such synchronization dates
back to at least the 17th century when Huygens observed synchronization
of two pendulums~\cite{roselblum_phys1996}. Synchronization phenomenon is
the subject of research efforts in a wide variety of fields. Large-scale
synchronization of firefly flashings and rhythmic applause in a large crowd
of audiences are but two examples in the general underlying principles beyond
mechanical objects.}
While many factors (\eg amplitude, stochastic noise) will
influence the phase of each oscillator, the following formula is a simplified
steady-state description of phase relationship for $N$ oscillators:
\begin{equation}
\label{eqn:kuramoto_glass_oscillator}
\frac{d}{dt}\phi_{i}(t)=\sum^N_{j=1} J_{ij}sin\Big(\phi_j(t)-\phi_i(t)\Big)
\end{equation}
Note that this simplified model ignores certain elements (\eg diffusion due
to noise) and is thus an approximation of a more complicated reality. Given 
such a differential equation describing a dynamic system, it can be shown
that a Lyapunov function in the following form exists~\cite{wang2019oim}:
\begin{equation}
\label{eqn:Lyap_kura}
H\Big({\Phi(t)}\Big)=-\sum_{\mathclap{i<j}} J_{ij}cos\Big(\phi_j(t)-\phi_i(t)\Big)
\end{equation}
This means that the system will generally evolve along a trajectory that
minimizes the Lyapunov function~\cite{nonlinear_system_analysis}. As a result,  the system's
stable states represent good solutions that minimize the right hand side
of Eq.~\ref{eqn:Lyap_kura}.
On a closer inspection, we see the resemblance of Eq. \ref{eqn:Lyap_kura} 
and the Ising model (Eq. \ref{eqn:Ising}). Specifically, when all phases 
($\phi_i$) are either 0 or $\pi$, the two formulae are the same.\footnote{In fact, the 
formulation of Eq. \ref{eqn:Lyap_kura} is similar to the classic XY spin 
model (again ignoring external field): each spin can point to any direction 
along an ``XY" plane and thus
can be represented by a phase ($\phi_i$). Ising model is thus a special
case of the XY model. In other words, a system of coupled oscillators form
an ``XY machine" (not an Ising machine). An XY state can be quantized into an 
Ising state ($\phi_i=0,\pi$) in a number of different ways. For the sake
of this paper, let us simply imagine direct quantization which rounds the 
phase to the nearest multiple of $\pi$.}
A number of oscillator-based Ising machines have been recently
proposed~\cite{wang2019oim, 10.1145.3316781.3322473, chou_analog_2015, wang2017oscillatorbased}. 
All these examples use LC tank oscillators.
While this is a common practice for analog circuit designers and relatively
straightforward for discrete-element prototypes, the use of LC tanks introduce
non-trivial practical challenges in integrated circuit (IC) designs. The lack
of high quality inductors and the usually high area costs of incorporating
them are common challenges for integrated RF circuitry. These desktop Ising
machines are a significant improvement (at least in size) over room-sized
Ising machines. But, for genuine wide-spread applications, we believe a
clean-slate IC-focused design is a valuable direction to pursue. Needless
to say, we believe there will be significant cross-pollination of different
approaches and future practice may very well be a confluence of multiple
styles of Ising machines.

\subsection{Accelerated simulated annealing}

Finally, a set of chips have been designed to accelerate simulated
annealing~\cite{Kirkpatrick671} or a variant of the classic algorithm. These
chips are often described as having tens of thousands of spins~\cite{yamaoka_isscc2015, 
takemoto_isscc2019}. In these designs, the spins are \emph{virtual} in that
they are bits in memory and manipulated by an algorithm (simulated annealing).
These machines are specially built to accelerate that algorithm. Hence 
we refer to such a machine as an \emph{Accelerated Simulated Annealer}
or ASA for short.

These ASAs differ fundamentally from \emph{physical} Ising machines. In
a physical Ising machine, nature guides the spins to a preferable state
according to physical laws. Thus, the machine can achieve ultimate speed
and energy efficiency in principle -- though it is entirely possible that
a particular physics exploited is slow or energy-intensive to control; or
it may be expensive to enable the physics, such as in creating
the cryogenic environment required for quantum annealing.

\subsection{Design space}

\begin{figure}[htp]\centering
\includegraphics[width=0.6\textwidth]{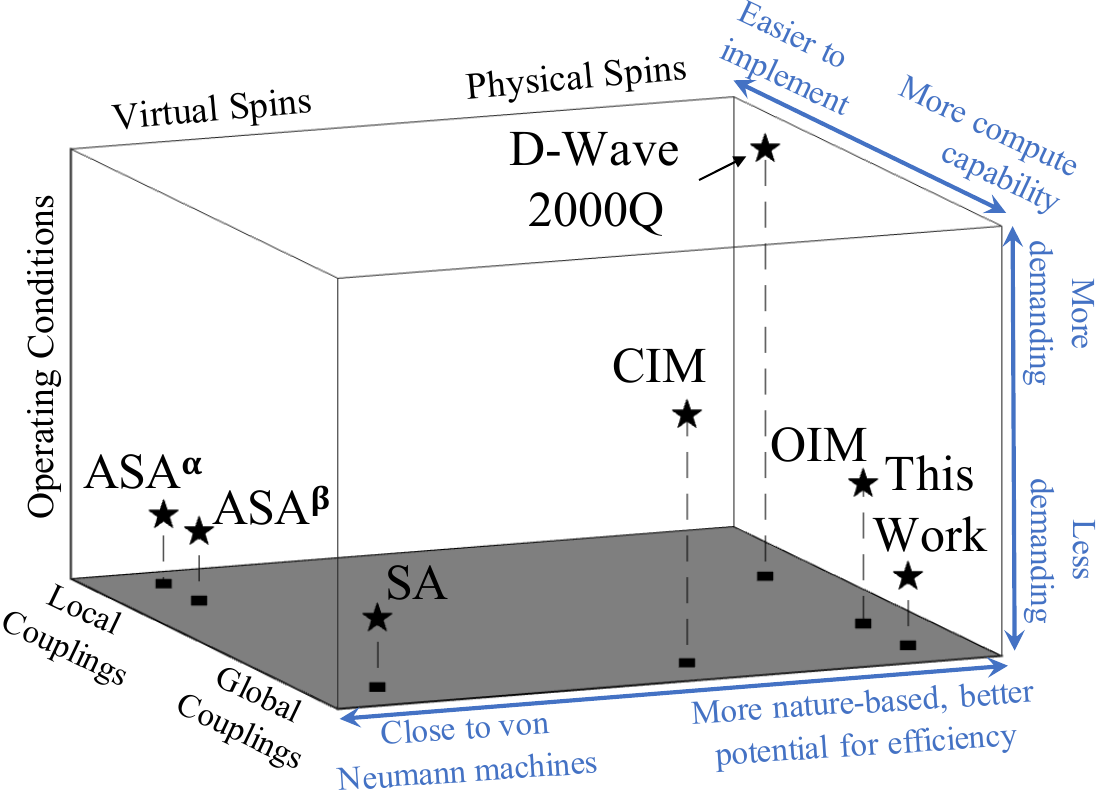}
\caption{High-level Ising machine design space showing qualitatively where
existing machines belong. OIM refers to a number of similar electronic
oscillator-based Ising machines~\cite{wang2019oim,8806780,9180748,chou_analog_2015,8993460}.
ASA$^\alpha$ and ASA$^\beta$ refer to machines in works
\cite{yamaoka_isscc2015} and \cite{takemoto_isscc2019} respectively.
}
\label{fig:taxonomy}
\end{figure}

Existing Ising machines can be placed in a design space along a number of
dimensions. \fig~\ref{fig:taxonomy} visualizes this high-level classification. 

The first dimension is whether they use physical or virtual spins. Note that
this aspect is more of a continuum than a binary distinction. In the case of
CIM, for instance, the interaction of the spins happens physically in the
fiber. But the appropriate amplitude of feedback is controlled using external
calculation.

A second important differentiator of Ising machines is the connection
topology. As discussed before, with a local connection, there is a limitation
on the kind of problems that can be mapped to the machine. When a problem
does not map directly, a transformation is needed to convert the problem,
usually by requiring many auxiliary nodes. The nominal number of spins a
machine provides is therefore a very poor representation of the machine's
capability. We will provide a quantitative analysis on this point in
Sec.~\ref{sec:evaluations}. At this point, we qualitatively place existing
Ising machines based on their connection topology on one axis in
\fig~\ref{fig:taxonomy}.


Finally, a third dimension characterizes how demanding the operating conditions
are. Nature-based computing is attractive for its potential in speed and energy
efficiency. But in some cases, the physics that allows nature-based computing requires
stringent operating conditions. D-Wave's quantum annealers are one such example.
Much of the system's bulk and energy consumption is to create the cryogenic operating
conditions for the system to work properly. Similarly, but to a lesser extent,
CIM requires stringent temperature stability and noise isolation to ensure the stability
of the interacting optical pulses. Electronic Ising machines, on the other hand,
can operate in room temperature and are unlikely to require a special supporting
environment. Nevertheless, similar to analog electronics, they will be more
sensitive to noise than conventional digital computers. Later, we will show 
some analysis of our design's behavior in a typical on-chip environment.

\section{Design of the CMOS IC Ising Machines}
\label{sec:proposal}

In this section, we start with a simplified system to provide some intuition 
about how common electronics can also make a physical Ising machine 
(Sec.~\ref{ssec:overview}); then describe in more detail the system architecture 
(Sec.~\ref{ssec:arch}) and the design of key circuits (Sec.~\ref{ssec:circuit});
and finally present theoretical analysis of the system's optimum-seeking 
behavior (Sec.~\ref{sec:theory}).

\subsection{Overview \& intuition}
\label{ssec:overview}

As already discussed before, existing physical Ising machine have different
strengths and weaknesses. The room-sized machines are vehicles for continued
scientific exploration of the underlying principles. It is particularly useful
to show the difference between ideal theoretical capabilities and what can be
achieved in practice. For instance, according to quantum adiabatic theory,
the system's Hamiltonian needs to be changed sufficiently slowly to guarantee
that the system stays in ground state.
Operated as such, the quantum annealer 
predictably provides sub-optimal solutions, as we will show later.

The question then becomes: can we build better (smaller, less power-intensive) 
physical Ising machines. And the answer is: yes, with electronics. In digital 
designs, electronic devices are often thought of as no more than the building 
blocks of functional units. But their behavior is also subject to physical 
laws that can be leveraged to perform nature-based computation. As it turns out, 
practical physical Ising machines can be built out of common devices such as 
capacitors and resistors. We start with one such simple design to show the 
working principle. Of course, this design is not yet a high-performance system. 
But as we will show later, with a proper architecture and careful design of key 
elements, a physical Ising machine built out of electronics is much more compelling 
than existing proposals. Additionally, it can be fabricated entirely in a CMOS 
process.\newline

\textbf{Intuition:} In the Ising model, when two nodes (say, $i$, and $j$) are 
strongly and positively coupled (\ie $J_{ij}$ is large and positive),  their 
spins are likely to be parallel ($\sigma_i = \sigma_j$). In this way, the term 
$-J_{ij}\sigma_i\sigma_j$ will contribute to lowering the energy. Conversely, 
a strong negative coupling ($J_{ij}$ is large and negative) will likely lead to 
anti-parallel spins ($\sigma_i = -\sigma_j$). Finally, weak coupling ($ |J_{ij}|$ 
is small) suggests that the two spins are more likely to be independent.

\begin{figure}[htp]\centering
    \begin{minipage}{.3\textwidth}\centering
    \includegraphics[width=\linewidth]{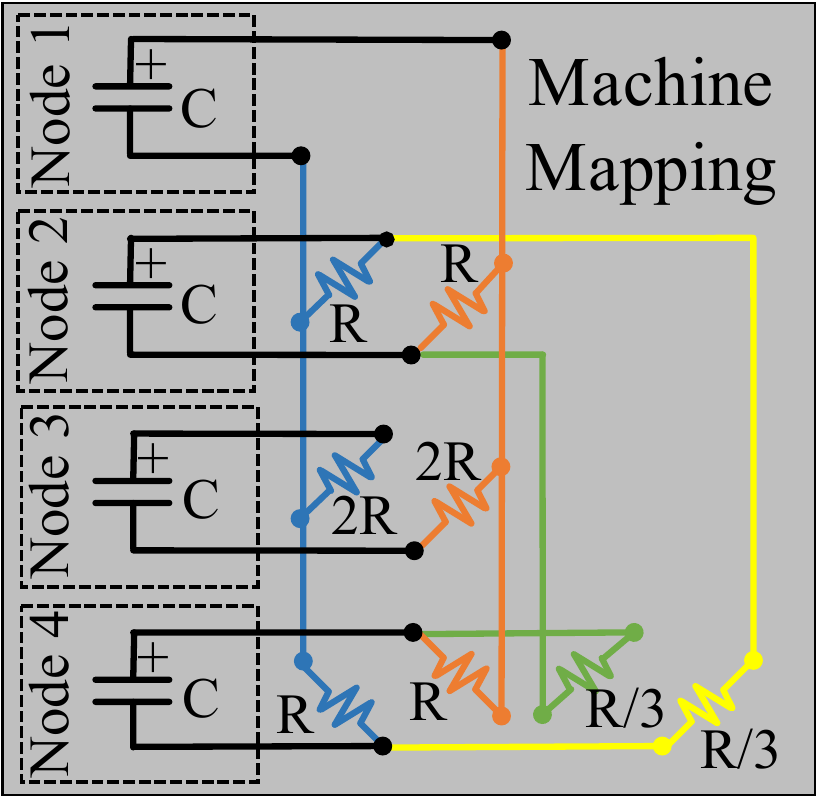}
    \end{minipage}
    \qquad
    \begin{minipage}{.32\textwidth}\centering
    \includegraphics[width=\linewidth]{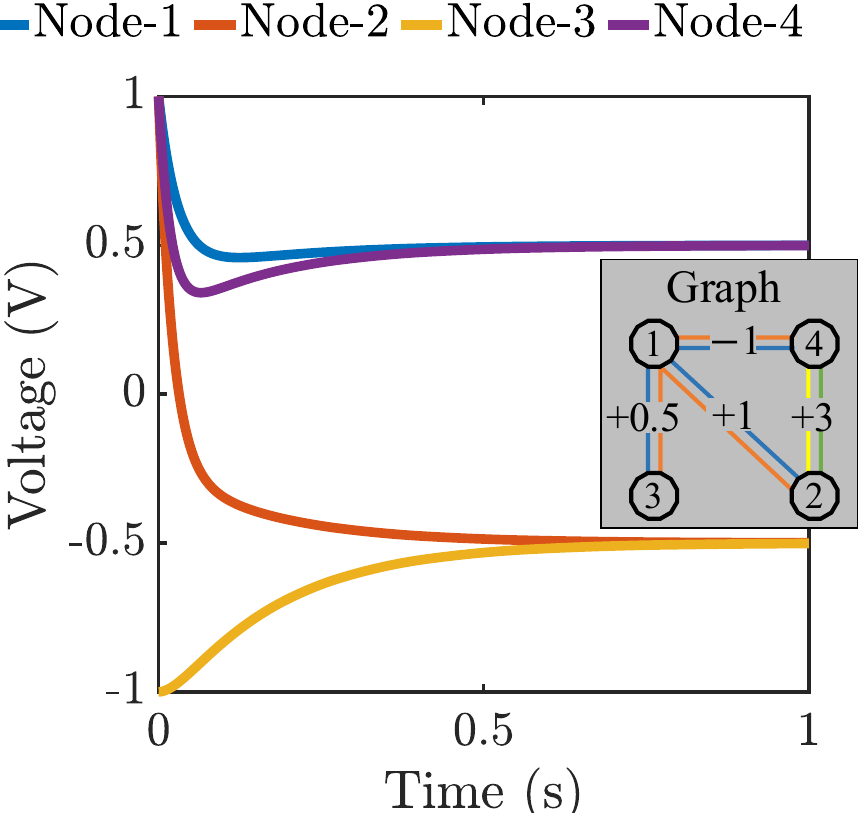}
    \end{minipage}
    \caption{Sample Max-Cut problem mapping to a 4-node resistive Ising machine 
    (left) and corresponding solution with Max-Cut value of $+4.5$ (right).}
    \label{fig:four_nodes}
\end{figure}

This behavior can be easily mimicked with resistively coupled capacitors. Consider 
representing a node with a capacitor, where the polarity of the voltage  across it 
represents the spin of the node. More specifically, in \fig~\ref{fig:four_nodes}, 
if a node has a spin of ``-1", the voltage at the upper terminal (top plate of the 
capacitor, labeled ``+") is lower than that of the lower terminal (bottom plate of 
the capacitor).  We can then connect nodes with different resistors. A strong coupling 
means high conductance (or low resistance), so that voltages of two nodes can more 
easily equilibrate. So, we set $R_{ij}\propto\frac{1}{J_{ij}}$. The sign of coupling can 
also be achieved by connecting either the same or opposite polarity terminals of the 
corresponding capacitors. \fig~\ref{fig:four_nodes} shows a simple 4-node system 
mapped from a logical graph of a Max-Cut problem with the labeled edge weights.
\footnote{Recall the coupling and edge weight relation, $J_{ij} = -W_{ij}$, in the 
Ising formula.} It is not difficult to see that the solution should separate the 
nodes into $\{1, 4\}$ and $\{2, 3\}$. Let's see how the machine functions.

The graph translates to couplings in a straightforward manner: 
$R_{ij}\!=\!\frac{R}{J_{ij}}\!=\!-\frac{R}{W_{ij}}$, where the sign indicates 
polarity of coupling. For instance, nodes 1 and 4 ($W_{1,4}=-1$) are parallel/positively 
coupled ($R_{1,4}\!=\!-\frac{R}{W_{1,4}}\!=\!R$). So a resistor of $R$ connects 
the upper terminals of nodes 1 and 4 and another connects the two lower terminals. 
Nodes 1 and 3 ($W_{1,3}\!=\!+0.5$) are antiparallel/negatively coupled ($R_{1,3}\!=\!-\frac{R}{W_{1,3}}\!=\!-2R$), 
so a $2R$ resistor connects the upper terminal of Node 1 and the lower terminal 
of Node 3, and another $2R$ resistor connects their remaining terminals.





Once initialized with random polarities, these coupled capacitors can indeed seek 
some equilibrium as shown in \fig~\ref{fig:four_nodes} (right). In this example,
the polarity of the capacitors at equilibrium indeed gives the best solution to 
the Max-Cut problem.

While this oversimplified design confirms the intuition that it \emph{can} find 
a solution, it is far from a robust design. For example, depending on the initial 
state and system scale, the voltages at equilibrium can be $0\,V$ or just too low 
for reliable readout. The equilibrium is also temporary because leakage will make 
all nodes decay to $0\,V$ eventually, rather than staying at the desired  voltage 
levels. Nevertheless, the resistively-coupled capacitor network is at the core of 
our proposed Ising machine. To induce and maintain the nodes at equilibrium, we can 
introduce a local feedback unit to make the node voltages bistable. For brevity, we 
will refer to such a Bistable, Resistively-coupled Ising Machine as BRIM.

\subsection{Architecture of an integrated design}
\label{ssec:arch}

We now discuss the architecture of a more complete system designed for integrated 
circuits. The system is illustrated in \fig~\ref{fig:BRIM_arch} and consists of: 
\ding{172} bistable nodes (N1 to N4) and their digital interface (DFF 1-4); 
\ding{173} coupling units connecting the nodes ($CU_{ij}$); \ding{174} programming 
units to configure coupling weights (including: memory, multiplexers, digital-to-analog 
converters, and the column control systems); and \ding{175} annealing control. We 
discuss each in turn as follows: 

\begin{figure}[htp]
    \centerline{\includegraphics[width=0.6\textwidth]{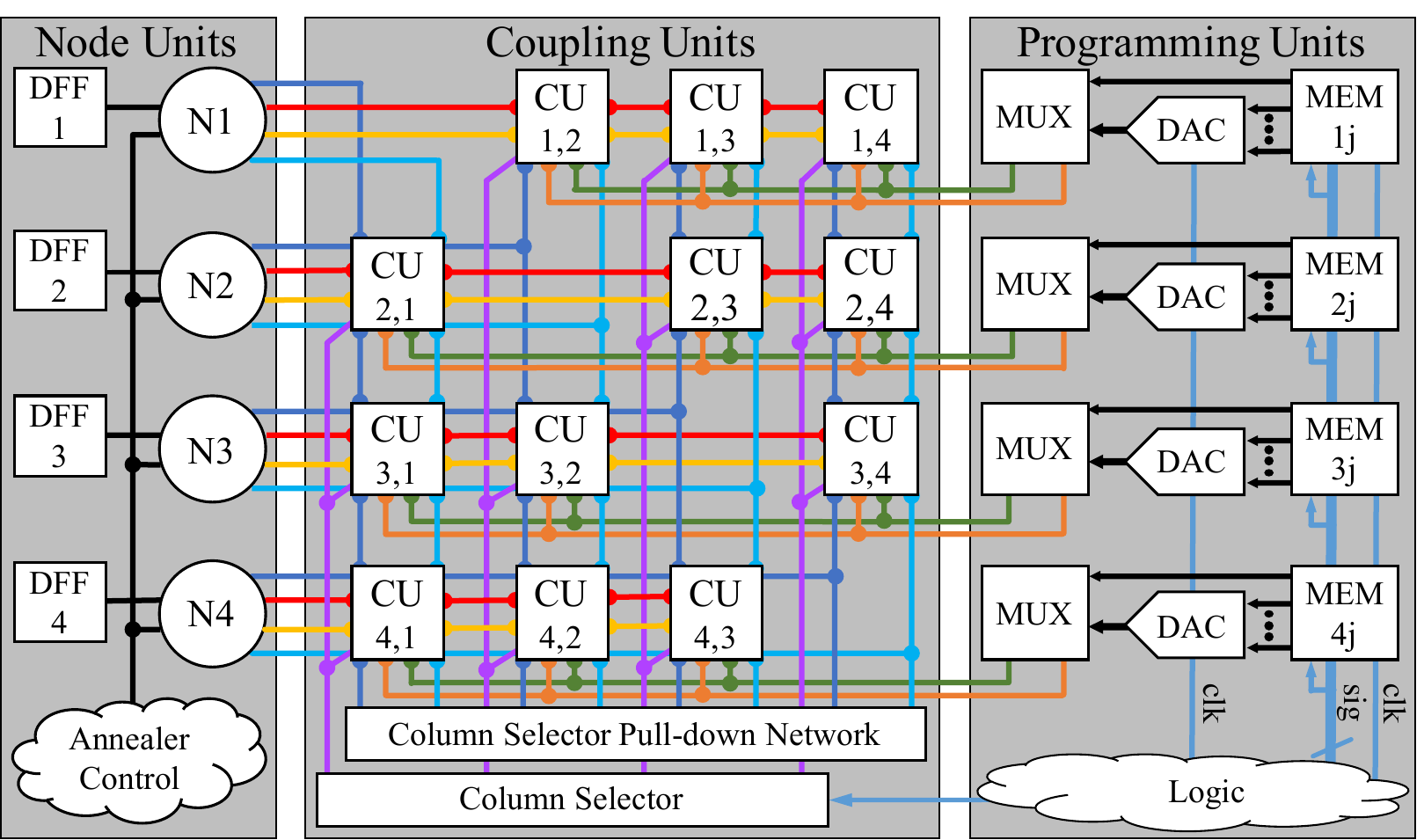}}
    \caption{Block diagram of BRIM system components. Nodes are $N_i$ and coupling units 
    $CU_{ij}$. The multi-colored interconnecting links are wires.}
    \label{fig:BRIM_arch}
\end{figure}

\subsubsection{\bf Nodes}

\begin{figure}[htp]\centering
    \begin{minipage}{.45\textwidth}\centering
    \includegraphics[width=\linewidth]{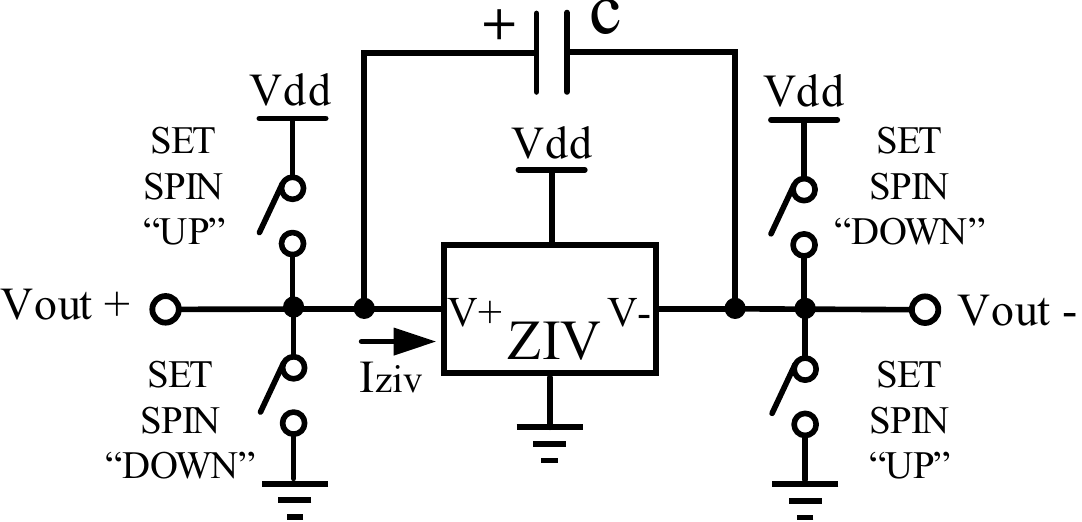}
    \caption{Simplified schematic of one BRIM node.\label{fig:node_schematic}}
    \end{minipage}
    \qquad
    \begin{minipage}{.32\textwidth}\centering
    \includegraphics[width=\linewidth]{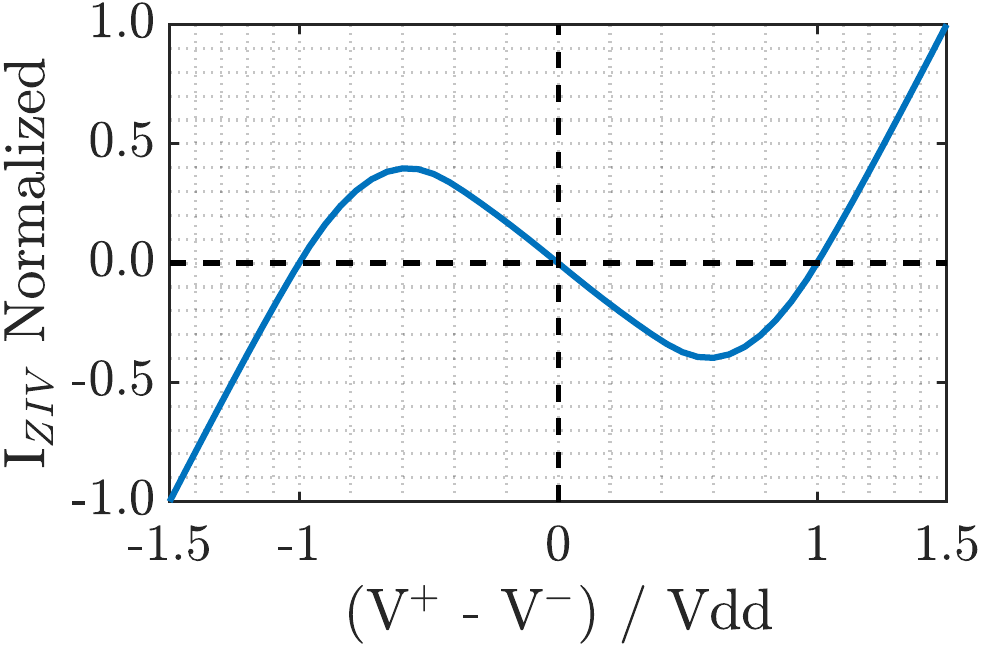}
    \caption{ZIV diode's IV characteristic.\label{fig:node_and_iv}}
    \end{minipage}
\end{figure}

The bistable nodes $N_i$ are shown at the left side of \fig~\ref{fig:BRIM_arch}. 
In \fig~\ref{fig:node_schematic}, we show a more detailed illustration of one node.
Recall that in the simplified circuit (\fig~\ref{fig:four_nodes}) the capacitor's 
voltage can be too low (even zero) as compared to electronic noise level to reliably 
indicate the node's spin. A feedback circuit is therefore needed to stabilize the 
voltages at the desired levels (\eg $\pm V_{dd}$). Two conditions are required: 
\ding{172} the capacitor should be charged according to its polarity when the voltage 
is between $-V_{dd}$ and $+V_{dd}$, and discharged when the voltage exceeds this range; 
and \ding{173} at low voltages, the feedback circuit should supply a low current in 
order not to overwhelm signals coupled from other nodes. 

Combining these considerations, we can design a feedback circuit with the current-voltage
(IV) curve as shown in \fig~\ref{fig:node_and_iv}. Because of the slanted ``Z" shape of 
the IV curve, we call such sub-system a ZIV diode. As seen in the IV curve, for capacitor 
voltages between $-V_{dd}$ and $V_{dd}$, the ZIV diode acts as an active element charging 
its voltage closer to $\pm V_{dd}$. Conversely, for voltages outside the range, the ZIV 
diode acts essentially as a linear resistor to discharge the capacitor. 
Each node is supplied with two pairs of switches to set the capacitor voltage to a known
spin state. This is useful for specific initialization or to change the system state as 
described in the ``Annealing Control" (Sec.~\ref{ssec:anneal_control}). Finally, the spin 
of the node can be read out by a digital latch (\eg a D flip-flop) connected to the ZIV 
diode's $V+$ terminal. This converts the spin to a bit (1 or 0).

\subsubsection{\bf Coupling units}

With an array of BRIM nodes, we have the flexibility to couple them in any desired topology. 
As we will discuss quantitatively later, an all-to-all coupling is computationally more 
useful and is thus our choice.

In our system, the coupling is (uni)directional. There are two separate coupling units (CU)
connecting node $N_i$ to $N_j$: $CU_{ij}$ and $CU_{ji}$. This directional coupling is 
achieved by connecting the node capacitor through a voltage buffer before sending to the CU. 
The coupling coefficient 
of both directions are of course the same ($J_{ij} = J_{ji}$). In principle, a bidirectional 
coupling has similar effects and is a simpler design point. But, empirically, we found
directed coupling tends to produce better solution quality at the expense of area. A more 
detailed analysis as to the reason for this phenomenon is left as future work.

\begin{figure}[htp]\centering
    \centerline{\includegraphics[width=0.40\textwidth]{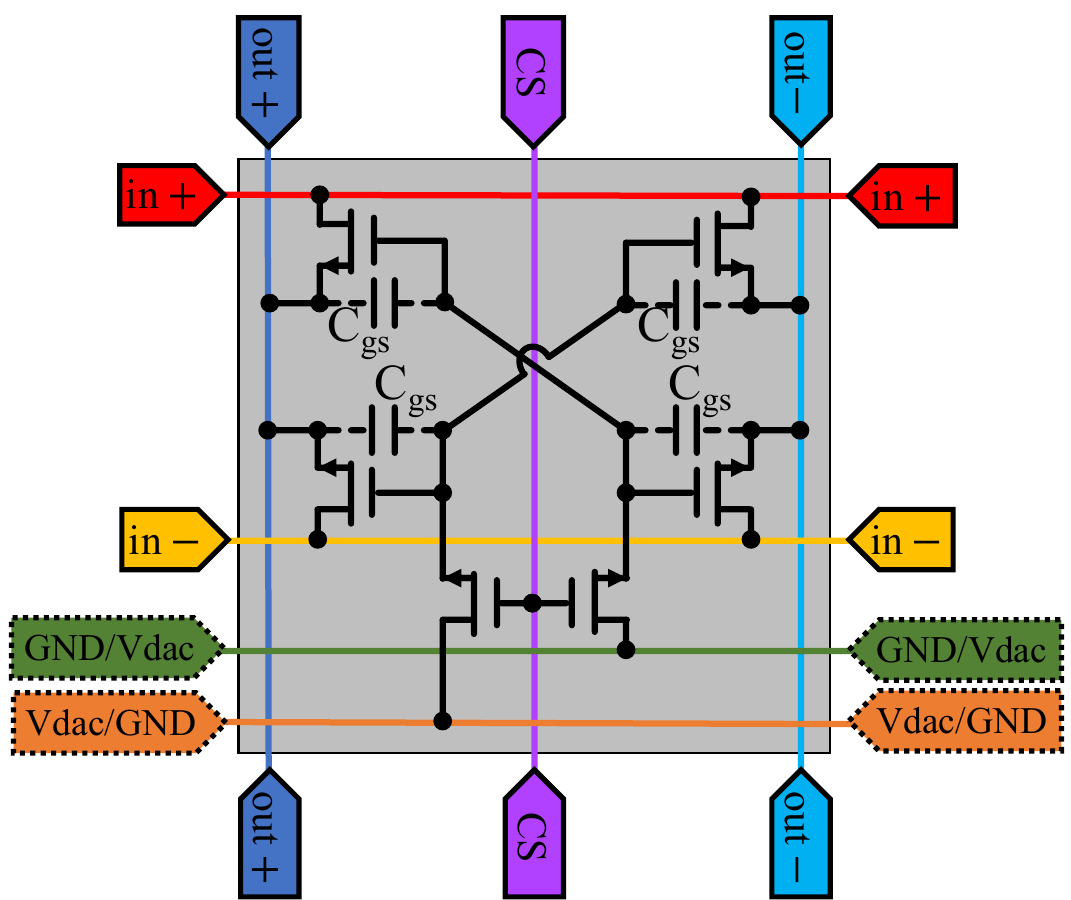}}
    \caption{Schematic of BRIM coupling unit. Color schemes are matched to
    system-level block diagram in \fig~\ref{fig:BRIM_arch}. CS is column selector.}
    \label{fig:coupling_unit}
\end{figure}

Because of the directed coupling scheme, each node has separate input and output terminals
(two each for fully-differential coupling), as shown in different colors in \fig~\ref{fig:BRIM_arch}. 
Similarly, each CU also has four input/output terminals, with one pair of transistors 
connecting the same polarity input/output nodes (parallel coupling) and another pair to 
establish anti-parallel coupling, as shown in \fig~\ref{fig:coupling_unit}.

The two pairs of transistors in the CU are biased in triode-region to act as variable/programmable
resistors. Depending on the polarity of the coupling coefficient ($J_{ij}$), only one pair 
of the transistors are turned on and biased to an appropriate non-zero $V_{GS}$. For example, 
to establish a negative (anti-parallel) coupling, the gates of the upper-right and lower-left
transistors are biased to a non-zero voltage ($V_{dac}$), while the other pair of transistors 
are biased at $0V$ (GND).

\subsubsection{\bf Programming units}
Both the initial nodal values and the coupling resistance are programmable. Programming of a 
resistor is achieved through a transistor with adjustable gate voltage. To the right of the coupling 
unit array in \fig~\ref{fig:BRIM_arch} is the programming array. This array consists of 
digital memory for storing the weights which drive an array of digital-to-analog converters 
(DACs). A small amount of such DACs are sufficient to program all the coupling units in a 
time-interleaved fashion per column. In such configuration, we need corresponding column 
selectors and pull-down logic as shown below the coupling units in \fig~\ref{fig:BRIM_arch}.

\subsubsection{\bf Annealing control}\label{ssec:anneal_control}

With the basic network of coupled bistable nodes, the system can
reach the local optimum determined by the initial state. Two commonly
used mechanisms for annealers are incorporated to allow the system 
to escape local
optima. First, the coupling strength is globally increased over time. 
In this way, at the very beginning, the machine is only weakly coupled, 
rendering the energy landscape relatively flat. This helps the system
explore the landscape in a coarse granularity. We choose an
exponential annealing schedule because (a) it is a common practice, and (b) it can be
conveniently achieved by discharging an appropriately sized 
capacitor as the global annealing
scheduler. The voltage from this capacitor is then used to control
the variable gain buffer in each node to achieve the change in coupling
strength.

Second, we also adopt a similar strategy as that used in simulated annealing. 
By performing a ``spin flip" of select nodes (\ie to change the spin to its
opposite value),
we can enter a neighboring state in the global phase space. This allows
the system to escape the current basin of attraction and
explore new regions. In simulated annealing, the probability of such bit flips
is a function of both energy difference due to the bit flip and the current
temperature. For implementation convenience, we only use the temperature
to decide the probability/frequency of spin flips. The temperature
follows the same exponential annealing schedule discussed above. In other
words, the frequency of spin flips decays exponentially. When a
spin flip is decided, we randomly select a node and use the switches shown
in \fig~\ref{fig:node_and_iv} to achieve the purpose.

With these supports,
our BRIM is used similarly to other Ising machines: first, program the
weights; then, select the annealing time; and finally, read out the state of
the nodes after the completion of the annealing schedule. 
Note that the system can be used in a number of
different ways: the annealing time can be adjusted; the spin flip 
frequency  can be tuned; 
the machine can also be used together with a
software-based search algorithm (\eg simulated annealing), perhaps by
searching a subspace.

\subsection{Integrated circuit design}\label{ssec:circuit}

\fig~\ref{fig:balanced_BRIM} shows the circuit of a BRIM node.
The design is fully balanced, and can therefore be understood 
by analyzing either half. The circuit can be understood as the following 
functional blocks:

\begin{figure}[htbp]
    \centerline{\includegraphics[width=0.7\textwidth]{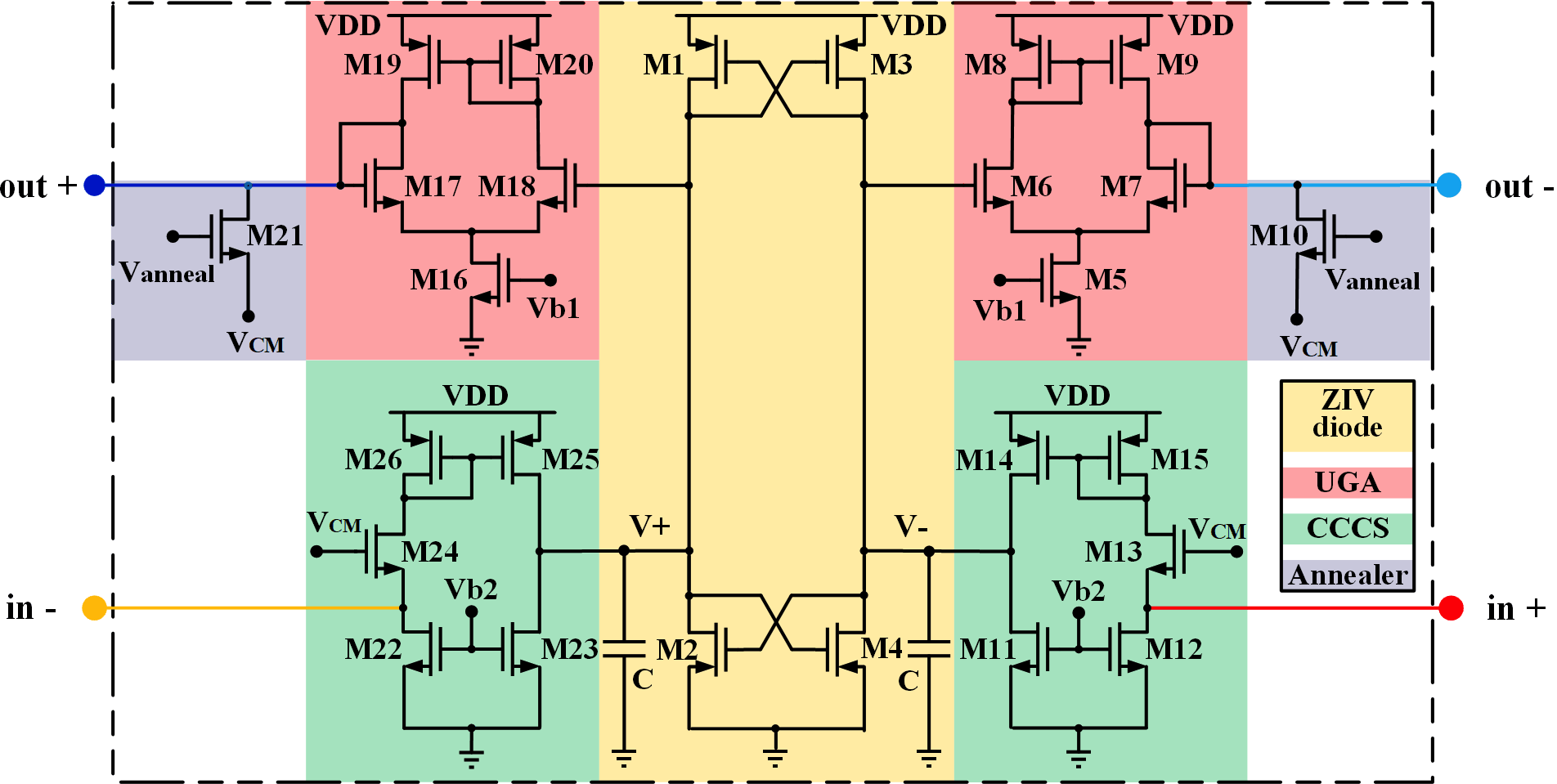}}
    \caption{High-level CMOS design of BRIM node.}
    \label{fig:balanced_BRIM}
\end{figure}

\begin{enumerate}

\item {Bi-stable node:}
Transistors $M_1$ to $M_4$ (center) are an integrated realization of the 
ZIV diode. It captures the overall IV characteristic shown in \fig~\ref{fig:node_and_iv} 
with a very simple circuitry, namely, two cross-coupled inverters. Together 
with capacitors $C$ connected on both sides of the ZIV diode, the core of 
the node with a bi-stable state is formed. The rest of the node is the 
interface to the coupling units.

\item{Output buffer:}
Recall that the voltage of the node goes through a buffer before being
connected to the coupling units. This buffer is implemented by transistors 
$M_5$ to $M_9$ (top right) which form a single-ended differential
amplifier configured in what is called a unity-gain topology.
The gain of the buffer changes with annealing -- as
we will show later -- making it a Unity Gain Amplifier (UGA).

\item{Input current control:}
The current from coupling units goes through a current-controlled
current source (CCCS) to connect to the nodal capacitor. Transistors 
$M_{11}$ to $M_{15}$ (bottom right) compose an integrated CCCS, which 
is modified from an active cascode summing circuit~\cite{comer2003high}.

\item{Annealing control:} $M_{10}$ operates as a variable resistor when
applying an annealing voltage $V_{anneal}$ to its gate. $V_{anneal}$ is a
global control signal supplied to all nodes according to a specific annealing 
schedule (\eg exponential decay). At the onset of the annealing process, $V_{anneal}$ 
is set to $V_{dd}$ which effectively shorts the output from the UGA setting its gain 
to $0\,V/V$ and disconnecting the node from the coupling network. As $V_{anneal}$ 
continues to decrease following an annealing schedule, the channel resistance of 
$M_{10}$ increases resulting in a reduced loading to the UGA and a gradual 
increase in its gain. At the completion of the annealing schedule, the $M_{10}$ 
is fully off, the UGA's gain is at its maximum value (about $0.9~V/V$ in our setup) 
and the node is fully connected to the coupling network.

\item{Coupling resistors:}
The coupling resistors (from \fig~\ref{fig:coupling_unit}) are implemented 
as bootstrap switches/transistors commonly used in low-voltage switched-capacitor 
analog circuits. The parasitic capacitance $C_{gs}$ of the bootstrap transistors 
is used to store and maintain the biasing voltage $V_{gs}$ set by the low-level 
DAC during programming. This voltage regulates the channel resistance of the 
transistors.

\end{enumerate}

\subsection{Theoretical analysis}\label{sec:theory}

\begin{figure}[htp]\centering
    \begin{minipage}{.35\textwidth}\centering
    \includegraphics[width=\linewidth]{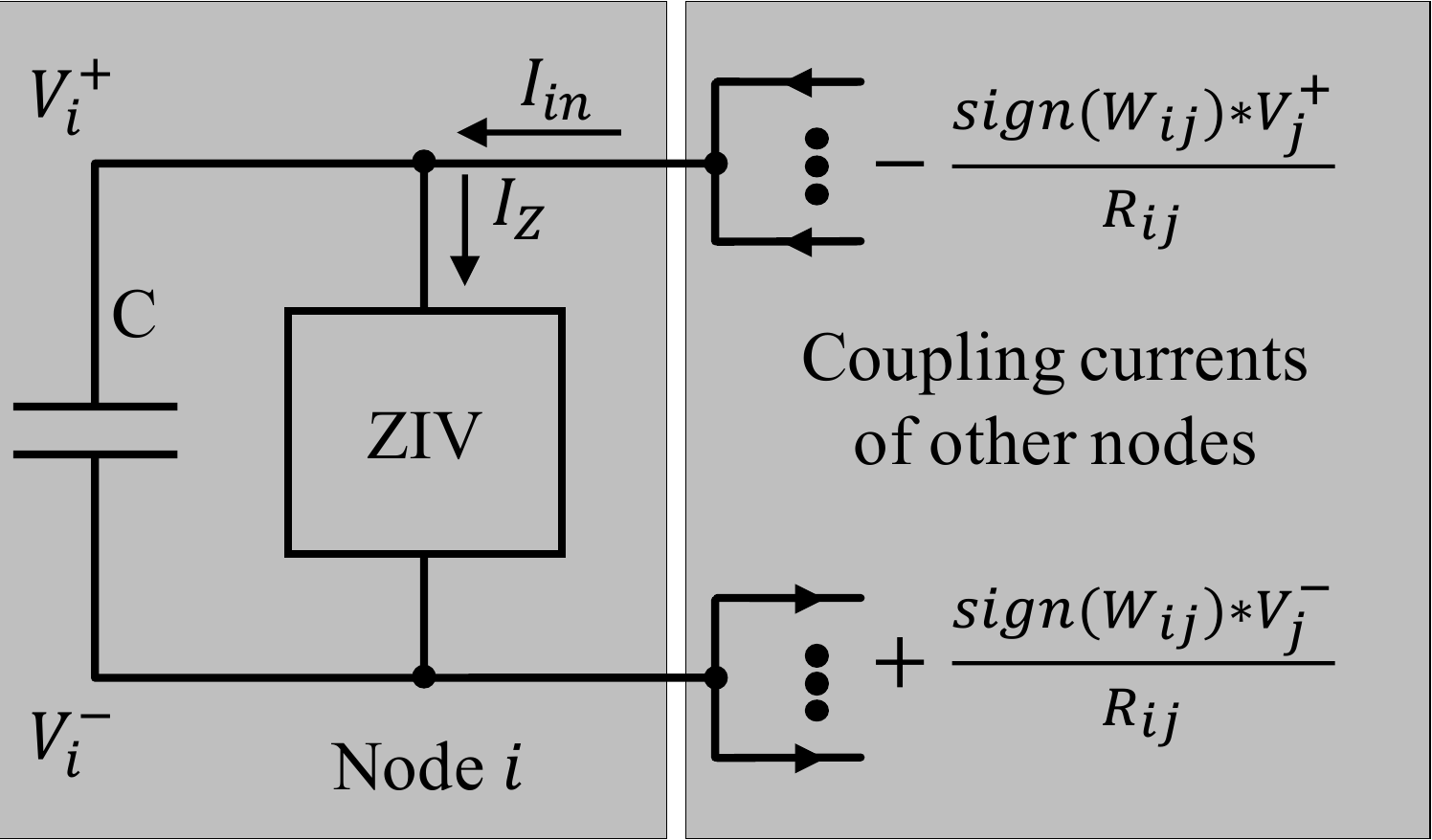}
    \caption{Simplified schematic of BRIM node.\label{fig:brim_node}}
    \end{minipage}
    \qquad
    \begin{minipage}{.3\textwidth}\centering
    \includegraphics[width=\linewidth]{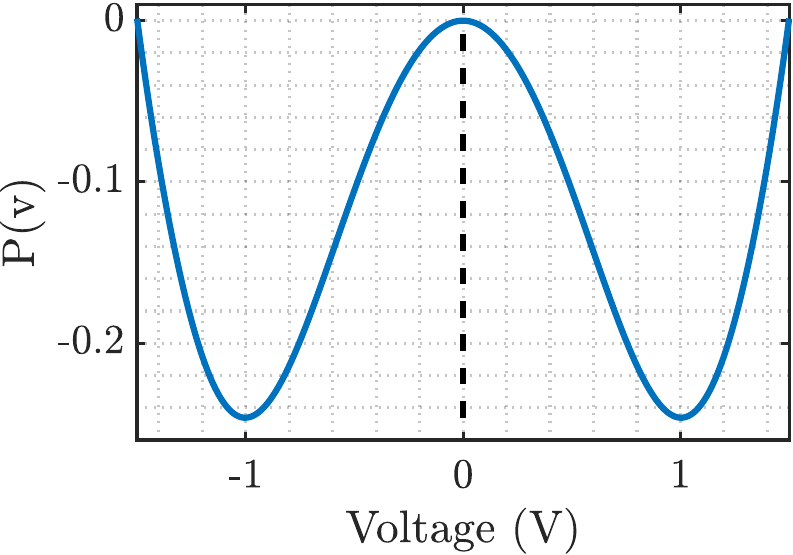}
    \caption{ZIV diode's IV characteristic.\label{fig:brim_well}}
    \end{minipage}
\end{figure}

It is often desirable to show some theoretical foundation for nature-based
computing machines. We present a Lyapunov stability analysis of BRIM to 
show its ability for optimum-seeking. \fig~\ref{fig:brim_node} shows a simplified 
node $i$ in BRIM where other nodes are coupled into $i$ from the right.
Writing the current equation, we get Eq.~\ref{eqn:rim_node_simplifed}, where $J_{ij}$ is the 
effective conductance of the coupling taking into account the polarity of
the coupling:
\begin{equation}\label{eqn:rim_node_simplifed}
    \dv{v_i}{t} = \frac{1}{C}(I_{in} - I_{ZIV}) =  \frac{1}{C} \left[ \sum_{j\neq i}J_{ij}v_j - g_{\!_{ZIV}}(v_{i}) \right]
\end{equation}

BRIM is a continuous time system where the state is summarized by
$v(t)=[v_1(t), v_2(t),...v_N(t)]^T$. (For brevity, in the following, 
$v_i(t)$ will be abbreviated as $v_i$.) Recall that in Lyapunov analysis, 
if a function $H(v)$ can be found such that $\dv{H(v)}{t}=0$ at point $v^e$
and $\dv{H(v)}{t}<0$ in the region around $v^e$, then $v^e$ is the 
equilibrium point~\cite{nonlinear_system_analysis}. In our case, this point 
is the solution to the equation set $\dv{v_i}{t} = 0;\ i=1..N$.

In other words, once the system enters a region, it will inevitably evolve
towards lowering $H(v)$ (because its time derivative is negative) until
it reaches the equilibrium point $v^e$.

To ensure $\dv{H(v)}{t}$ is non-positive, we can construct
it to be in a negative square form. This can be achieved 
by imposing the following construction rule.
\begin{equation}
\label{eqn:rule}
\pdv{H(v)}{v_i}=-\alpha\dv{v_i}{t}; \ \alpha>0
\end{equation} 

Following the chain rule, it is not difficult to see that:
\begin{equation}
\label{eqn:H_partials}
\dv{H(v)}{t} =\sum_{i}\left(\pdv{H(v)}{v_i}\dv{v_i}{t} \right)
=-\alpha\sum_{i}\left(\dv{v_i}{t}\right)^2
\end{equation}

One choice of a Lyapunov function satisfying the conditions in Eq.
\ref{eqn:rule} is shown in Eq.~\ref{eqn:Lyap_RIM}, where $P(v_i)$ is obtained
from $g_{\!_{ZIV}}(v_i)$ by integration over $v_i$. 
\begin{equation}
\label{eqn:Lyap_RIM}
H(v)= \frac{\alpha}{C}\Big(-\sum_{i<j} J_{ij}v_jv_i + \sum_i P(v_i)\Big)
\end{equation}

It is important to notice that the ``Z" shape of $g_{\!_{ZIV}}(v)$ will give 
$P(v)$ a double-well profile (as shown in \fig~\ref{fig:brim_well}) with two 
stable equilibrium points at voltages corresponding to two non-trivial 
zero-crossings of the IV curve (\eg $v_i= \pm 1V $) and a saddle point at 
$v_i=0V$. Given the double-well profile of $P(v)$, the state of a stable 
solution will consist of voltages at (or at least very close to) these 
equilibria (\eg $\pm 1V$). Thus, the second term of Eq.~\ref{eqn:Lyap_RIM} 
will be (close to) a constant. Consequently, minimizing $H(v)$ is equal to 
minimizing the first term $-\sum_{i<j}J_{ij}v_iv_j$, which is the Ising formula.

Note that this analysis does not guarantee that the system will converge to a
\emph{global} minimum as it depends on the energy landscape and initial 
conditions. This is similar to other annealers: None has strong guarantees 
for reaching ground state in a typical usage scenario (as opposed to ideal 
conditions) or in an efficient manner. For example, it is shown that simulated 
annealing can reach ground state in a system with a finite phase space. However, 
the time it takes to do so may be longer than enumerating the space~\cite{varty2017simulated}.

\section{Experimental Analysis}
\label{sec:evaluations}

In this section, we provide some experimental analysis of BRIM by

\begin{enumerate}
\item Comparing Ising machines at a high level (Sec.~\ref{ssec:highlevel});
\item Discussing architectural design choices and various sensitivity issues 
(Sec.~\ref{ssec:detail}); and
\item Showcasing a discrete-component prototype (Sec.~\ref{sec:BRIM prototype}).
\end{enumerate}

First, we describe the machines (Sec.~\ref{ssec:machines}), the benchmarks 
(Sec.~\ref{ssec:benchmarks}) used in the analyses, and sample BRIM modeling (Sec.~\ref{ssec:sample_model}).

\subsection{Ising machines}
\label{ssec:machines}

We compare a simulation based model of BRIM to 4 other machines using both 
physical and virtual spins. We use results reported in literature when 
direct measurement or modeling is unavailable.

\begin{enumerate}
\item \textbf{D-Wave:} we use 2000Q which is the latest quantum annealer
\cite{PhysRevB.82.024511} at the time of this documentation. We run jobs 
using the API provided by D-Wave~\cite{dwavesys}. For each graph problem, 
we collect 50 samples.~\footnote{In terms of timing, we do not specify 
any constraints, and adopt D-Wave default values. Specifically: $20\,\mu s$, 
$198\,\mu s$, $21\,\mu s$, and $11.7\,ms$ respectively for annealing, data 
readout, inter-sample delay, and programming time~\cite{dwave_timing}.} 
\item \textbf{CIM:} is an optical Coherent Ising Machine~\cite{inagaki2016cim}. 
There is no known public access to the actual hardware and no model available 
for simulation. Thankfully, there are reported results for two commonly used 
benchmarks and a 2000-node complete graph that allow us to make meaningful 
comparisons. 
\item \textbf{OIM:} is an electronic Oscillator-based Ising Machine
\cite{wang2019oim}. For simulation, we use the Kuramoto MATLAB model-based
code provided in~\cite{wang2019oim,wang2017oscillatorbased}.
We perform configuration sweep to obtain best results.
\item \textbf{ASA:} refers to a number of related designs of Accelerated 
Simulated Annealers~\cite{yamaoka_isscc2015, takemoto_isscc2019}. These 
accelerators use virtual spins and are straightforward to model based on 
the description in literature. We focus on one with 30,000 nominal spins. 
In this design, the coupling follows a near-neighbor pattern dubbed the 
King's graph. All nodes are grouped into 4 groups. For every annealing step 
($0.22\,\mu s$), nodes in one of the groups will process in parallel: 
they read off neighbors' spins and the associated weights to compute 
whether keeping the same spin or inverting its current spin provides a 
lower energy in the neighborhood. In addition, random bit flips similar 
to those in standard simulated annealing algorithms are also adopted.
\end{enumerate} 
As explained in Sec.~\ref{sec:background} (box 3 of \fig~\ref{fig:ising_machine}),
machines without all-to-all coupling (D-Wave and ASA) require preprocessing 
of input graph through minor embedding~\cite{cai2014practical}. This 
process currently takes significant time on a conventional computer 
(more details in Sec.~\ref{ssec:detail}) and can often fail for larger 
graphs. ASA uses the King's graph which is even more limited than the 
Chimera graph in D-Wave, the minor embedding process takes even longer 
time to complete and  results in more physical nodes needed. In our 
high-level analysis, we ignore the significant time needed for this 
preprocessing.  

Finally, we also added an optimized version of the Simulated Annealing 
(\textbf{SA})~\cite{Isakov_SA} algorithm as a reference written in C++. 
The execution time is measured on server cluster nodes of Intel Xeon 
Platinum 8268 CPUs at 2.9GHz with 371GB of RAM.

\subsection{Benchmarks}\label{ssec:benchmarks}

To compare these systems, we use a set of popular graphs called ``Gset''
(and their derivatives) with diverse node sizes and edge densities. 
These graphs are just weighted, undirected graphs and not associated 
with any specific problem. They naturally correspond to an Ising formula. 
Thus, we are comparing different machines optimizing the same set of 
Ising formulae.

Because of the direct mapping between an Ising formula and the Max-Cut 
problem, and that many algorithms have been developed for optimizing 
Max-Cut, it is convenient for researchers, especially Ising machine 
designers, to \emph{interpret} these graphs as specifying a Max-Cut 
problem, which we follow in this paper. Note that this does not mean 
we can only solve a Max-Cut problem as already discussed in 
Sec.~\ref{sec:background}.

The graphs we use can be divided into the following groups:
\begin{enumerate}

\item{\emph{Regular graphs}:} We use the original Gset graphs from Stanford
\cite{YinyuYe_Gsets}. These graphs have between 800 and 20,000 nodes. 
The edges as well as the weights of such edges, were generated 
probabilistically, sometimes between +1 and -1, and sometimes all +1. 
We only use those graphs with less than 2000 nodes in our experiments.

\donotshow{
\item{\emph{Small graphs}:} Although supporting nominally 2048 spins, D-Wave's 
machine can not map even the smallest graph in Gset. We therefore 
generate graphs with smaller node sizes (\eg 120) and/or edge densities
so that they can be successfully mapped onto D-Wave. For this purpose, we 
used "rudy", a machine-independent graph generator~\cite{rudy_generator}, 
which is the same generator used to produce the Gset graph suite.
}

\item{\emph{Tiny graphs}:}\label{subsubsec:tiny_graphs}
Finally, we also generated fully-connected graphs with random edge 
weights, and node sizes ranging from 16 to 32 (in increments of 4). 
Each node size has 20 sample graphs, for a total of 100 graphs. For 
these graphs, we are able to enumerate all possible spin combinations 
to determine the actual maximum cut.
\end{enumerate}

\subsection{BRIM sample modeling}\label{ssec:sample_model}

\begin{figure}[htbp]\centering
    \includegraphics[width=0.5\textwidth]{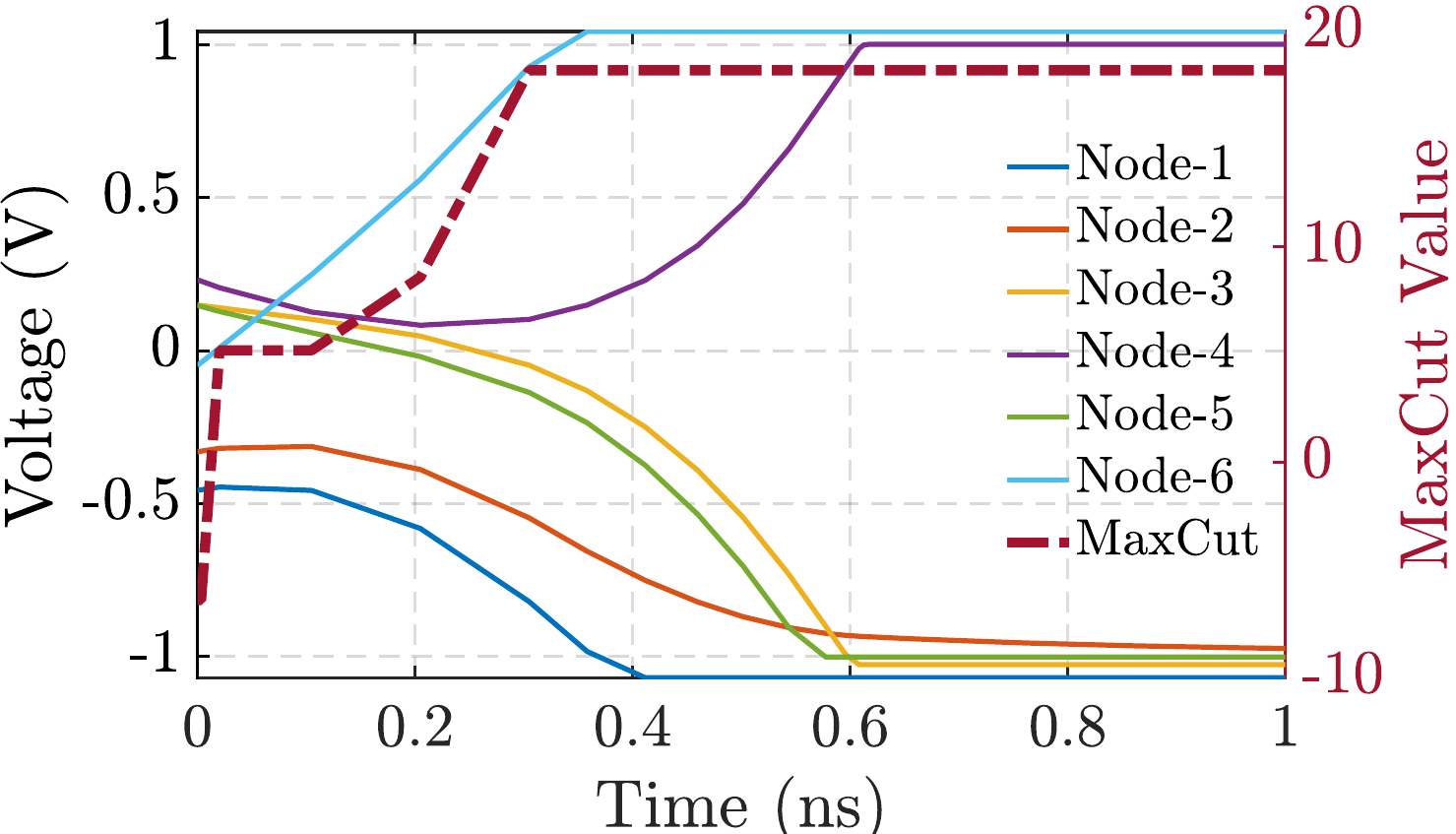}
    \caption{Sample state of BRIM machine solving a 6-node David star 
    problem. Right axis shows magnitude of MaxCut value over time 
    corresponding to the dashed line. See graph problem illustration 
    in \fig~\ref{fig:discrete}.}
    \label{fig:brim_baseline}
\end{figure}

We demonstrate an example model of the continuous-time BRIM system before 
getting to detailed analysis of the diverse machines. In this example, we 
use MATLAB's nonstiff, single step, 5th-order differential solver (ode45).
Listing~\ref{lst:brim_baseline} shows the sample BRIM implementation of 
Eq.~\ref{eqn:rim_node_simplifed}. In listing~\ref{lst:brim_base}, we 
demonstrate the setup and test for solving a simple 6-node David-graph 
problem the sample BRIM machine. \fig~\ref{fig:brim_baseline} shows a 
sample voltage transition of each node in the BRIM system.

\begin{center}\begin{tabular}{c}
\begin{lstlisting}[caption={BRIM sample (without annealing or spin-flip)},label={lst:brim_baseline},linewidth=8.5cm]
function dvdt = brim(~, v, W)
    C   = 49e-15;                          % Node capacitor
    Rc  = 31e3;                            % Coupling resistor
    ziv = @(v)(-2.1563e-05 .* v.^5) + ...  % Diode
              (  1.017e-04 .* v.^3) + ...  % ...
              ( -2.231e-05 .* v);          % ...

    J   = -W ./ Rc;                        % Ising transformation
    
    dvdt = ((J * v) - ziv(v)) ./ C;        % BRIM expression

    dvdt((v < -1) & (dvdt < 0)) = 0;       % Model diode range
    dvdt((v >  1) & (dvdt > 0)) = 0;       % ...
end
\end{lstlisting}
\end{tabular}\end{center}

\begin{center}\begin{tabular}{c}
\begin{lstlisting}[caption={Baseline BRIM test code},label={lst:brim_base},linewidth=8.5cm]
function sample
    % David-graph problem with maximum cut value of 18.2
    src = [1  1  1 1  2.0  2 2.0 3  3.0 4  4  5.0];
    dst = [2  3  5 6  3.0  4 6.0 4  5.0 5  6  6.0];
    wgt = [1 -1 -2 6 -0.2 -2 4.4 4 -4.4 6 -1 -0.2];
    
    spin= length(unique([src dst]));
    W1  = sparse(src, dst, wgt, spin, spin);
    W2  = sparse(dst, src, wgt, spin, spin);
    W   = full(W1 + W2);

    % Solve BRIM system
    F   = @(t,v)brim(t, v, W);             % Work function
    T   = [0 1e-9];                        % Annealing time
    I   = rand(spin, 1) - 0.5;             % Initial states
    sol = ode45(F, T, I);                  % Solver
    
    % Display solution
    plot(sol.x, sol.y);
end
\end{lstlisting}
\end{tabular}\end{center}

    


\subsection{High-level comparison}\label{ssec:highlevel}

It is important to keep in mind that Ising machines are far from mature. 
Early designs and prototypes are necessarily experimental in nature and 
thus lack the polish of, say, a classical computer architecture. Much of the 
performance difference may be due to the art of prototyping rather than 
the fundamental science of the mechanism being exploited. This is perhaps 
especially the case for D-Wave, as in our comparison, it is the only 
actual hardware that we have access to. (CIM and OIM both have hardware 
prototypes but are not accessible to us.)

We start with a crude, high-level comparison of different Ising machines. 
There are several practical factors that make this comparison crude and 
incomplete. First, there is no single problem that can be measured on 
all machines. This is primarily because CIM only reported raw data on a 
very specific set of benchmarks and we are unaware of any reliable model 
of the physics that is publicly available. Second, the workload of 
optimization usually allows tradeoff between speed and quality of the 
solution. Ideally, we will fix one metric (say, execution time) and
compare the other (quality of solution). But in some cases, such control 
is unavailable to us. Third, the execution result depends on initial 
conditions. So any single run is subject to random chances. The common 
practice of using these machines is doing  multiple runs and taking the 
best solution, which we follow. But this value should still be regarded 
as a random variable. 

\subsubsection{Area and power estimates}\label{ssec:area_and_power}

\begin{figure}[htbp]\centering
    \includegraphics[width=0.32\textwidth]{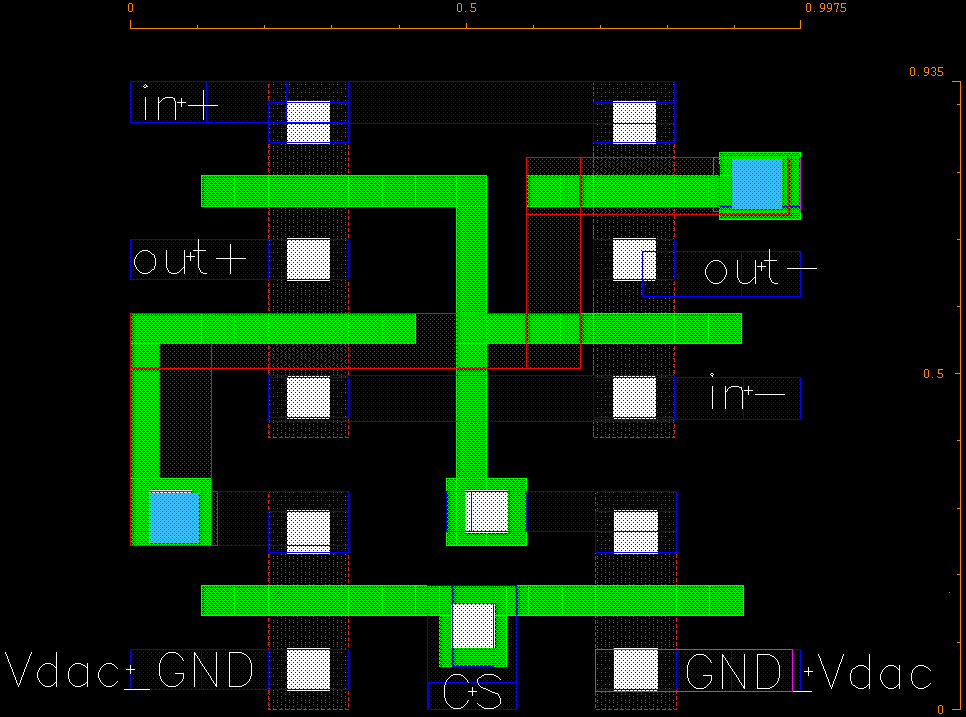}
    \caption{Preliminary design and layout of a coupling unit.}
    \label{fig:cu_layout}
\end{figure}

We use reported D-Wave and ASA power numbers from literature. Due to the
nature of the underlying physics being exploited, it is unlikely for 
these machines to reach chip-level size in the near future, if ever. 
For BRIM, we envision an integrated circuit implementation. For a rough
sense of the product, we perform a preliminary design and layout 
of components in Cadence using generic 45 nm PDK. For BRIM, the
area will be dominated by the 4 million coupling units (CU).
As can be seen in the layout shown in \fig~\ref{fig:cu_layout}, one coupling 
unit takes about $1 \mu m\times1\mu m$. Additionally, a BRIM
node has the size ($9.\,5\mu m \times 9\,\mu m$) and a DAC for the $\pm 1$ weights 
is ($2\,\mu m \times 1\,\mu m$). Combined, the rest of the system outside the
coupling network will be less than 1\% total area and even less in larger systems. 

Power consumption was collected in Cadence modeling a 6-node system 
(\fig~\ref{fig:discrete}) and measures $620\,\mu W$. Since the Gsets 
tested (Table~\ref{tab:comp}) are all sparse (with 16.4 edges/node 
ratio), we simply extrapolated to $200\,mW$ for 2000-nodes for a 
coarse grain, order-of-magnitude estimate. A simplified first-principle 
calculation shows the power to be $\approx 250\,mW$.

\subsubsection{Sample 2000-spin machine}

With these caveats in mind, Table~\ref{tab:comp} shows high-level metrics; 
the estimated power and chip area (when applicable), the execution time 
(machine annealing time), and solution quality of sample designs
with 2000-spins. The cut value itself changes a lot: 
some graphs have a cut value around 100, others around 2000.
A better measure of solution quality is thus not the absolute cut value,
but the \emph{distance} (how close the solution is)
to the best known cut value in literature.

\begin{table}[htp]
\caption{High-level comparison of diverse machines and metrics.
\label{tab:comp}}
\centering\footnotesize
\setlength{\tabcolsep}{1pt}

\begin{tabular}{|c|r|} \hline
\multicolumn{2}{|c|}{\bf{Parameters}}\\ \hline
\multicolumn{2}{|c|}{Power (W)}      \\ \hline
\multicolumn{2}{|c|}{Area ($mm^2$)}  \\ \hline
\multicolumn{2}{|c|}{Effective Spins}\\ \hline
\multirow{2}{*}{G22}    & Distance    \\
                        & Time        \\ \hline
\multirow{2}{*}{G39}    & Distance    \\
                        & Time        \\ \hline
\multirow{2}{*}{K2000}  & Distance    \\
                        & Time        \\ \hline
\end{tabular}
\;
\setlength{\tabcolsep}{2pt}
\begin{tabular}{|c|c|c|c|c|} \hline
\bf{CIM}    & \bf{OIM}      & \bf{ASA\scriptsize{$^U$}}  & \bf{SA}       & \bf{BRIM}         \\ [0.5ex] \hline
$210$       & $-$           & $\approx 1$   & $\approx 1$   & $\approx 250\,m$  \\ \hline
$-$         & $-$           & $\approx 24$  & $\approx 10$  & $\approx 5$       \\ \hline
$2000$      & $-$           & $160$         & unlimited     & $2000$            \\ \hline
$46$        & $47$          & $46$          & $45$          & $46$              \\
$5\,ms$     & $0.8\,ms$     & $70\,\mu s$   & $1.6\,ms$     & $0.25\,\mu s$     \\ \hline
$47$        & $47$          & $45$          & $47$          & $46$              \\
$5\,ms$     & $0.4\,ms$     & $70\,\mu s$   & $1.8\,ms$     & $0.25\,\mu s$     \\ \hline
$0$         & $-$           & $107$         & $122$         & $-4$              \\
$5\,ms$     & $-$           & $60\,ms$      & $65\,ms$      & $2.2\,\mu s$      \\ \hline
\end{tabular}


\end{table}

\begin{figure}[htp]\centering
     \includegraphics[width=0.55\textwidth]{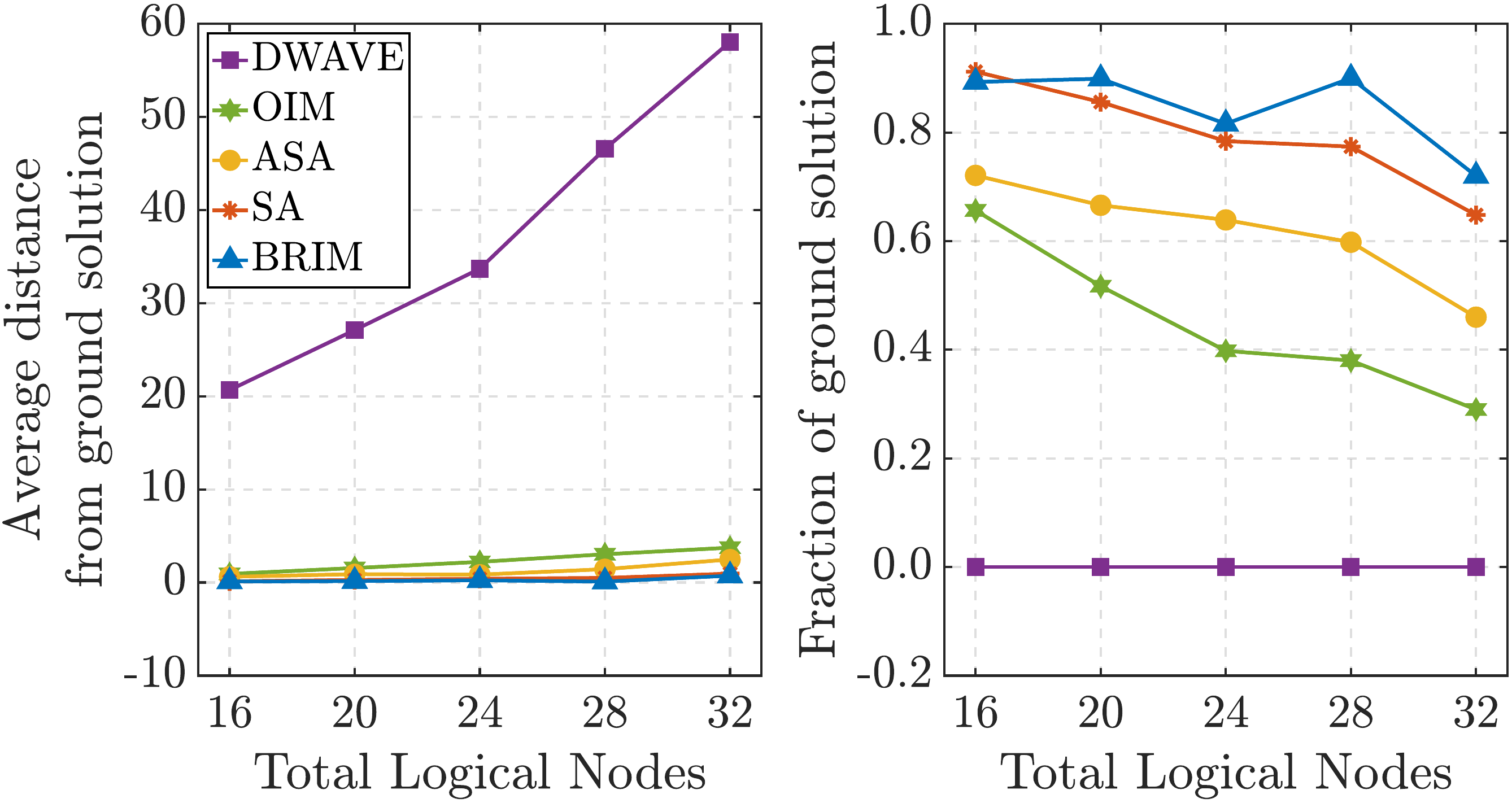}
    \caption{Tiny graph comparison on D-Wave. All systems have $20\,\mu s$ 
    annealing time.
    \textbf{Left}: Solution distance from known enumerated ground state solution.
    \textbf{Right}: Probabilities of going to ground state solution.}
    \label{fig:small_graphs}
\end{figure}

\begin{enumerate}
\item{\textbf{CIM comparision}}:
    We use CIM running graphs \emph{G22}, \emph{G39}, and \emph{K2000} as baseline 
    because these are tested on CIM and reported (\fig 3 of reference~\cite{inagaki2016cim}). 
    These graphs cannot be mapped on D-Wave. For other machines, we try to match CIM's 
    solution quality and show execution time. We see that BRIM can obtain similar quality 
    solutions 4 orders of magnitude faster with a power consumption about 3 orders of 
    magnitude better.
\item{\textbf{Tiny graph comparison}}:
    Recall that for these graphs we know the ground state solution from enumeration, 
    and we verify that BRIM, OIM, and ASA$^U$ can reach the ground state. Yet, even 
    for these tiny graphs, D-Wave does not reach the ground state, with the \emph{best}
    solution having an average distance of 20. A more detailed view of the solution 
    quality can be seen in \fig~\ref{fig:small_graphs} showing average solutions and
    probabilities of going to the ground state. SA and BRIM can both achieve ground 
    state often. But BRIM can do so with higher probability. These are concrete 
    examples where the theoretical ability to reach the ground state for an Ising 
    machine is just that: a theoretical ability.
\item{\textbf{To sum}}:
    We see that the room-sized machines (CIM and D-Wave) do not show any tangible 
    advantage in solving our weighted Max-Cut optimization problems. D-Wave can 
    only map the smaller problems due to the connection limits. Even on these 
    smaller problems, its solution quality trails behind others. Recall this 
    limited connection means additional compute time to perform minor embedding, 
    which at the moment takes $5\,s$, on average. CIM is less power-hungry and can map
    bigger problems due to its all-to-all connections. Nevertheless, there is no
    tangible advantage in any figure of merit. Again, these machines may (or may
    not) prove useful for scientific exploration and may (or may not) show some
    qualitative superiority at some other scale or at a future time.
    
    The most important take-away point is that just because the machine leverages
    nature to perform computation does not necessarily make it efficient. Much
    engineering diligence is needed to convert some theoretical possibilities to
    tangible practical benefits. 
\end{enumerate}

\donotshow{
The next two rows show small- and tiny-graph comparisons intended to contrast 
D-Wave to others.  Here, because D-Wave's current interface allows only fixed 
annealing time, we use the same  annealing time for all machines.  Recall that 
these are the best answers in 50 runs for each. We see that in these cases BRIM 
always achieves the best result among all machines, while D-Wave does not.
}

\subsubsection{Electronic designs}

Next, we look at the electronic designs.
While the current OIM prototype is a desktop machine,
in principle it can be scaled down in size and up in frequency. The primary
unknown is to what extent the nodes can scale down and still operate like
ideal LC-tanks. For this study, we assume a 100MHz frequency. 
Admittedly, this is nothing more than a rough guess. We see that OIM is perhaps 
comparable to BRIM, though subjectively, we feel that the practical challenges 
are far more daunting than in BRIM.

We also compare ASA with BRIM. ASA is essentially a specialized computer
doing algorithmic search for better cut values. Clearly, ASA takes advantage
of the tremendous cumulative improvements of CMOS technology to provide relatively
fast and efficient computation. However, it is still a modified von-Neumann 
machine (with its only program hardwired). In contrast, BRIM is a physical 
Ising machine where nature is performing the computation. Indeed, here, we 
see that to get similar quality solutions on small graphs, BRIM is still 
producing better results while consuming a tenth of the power, with a 5x 
lower area cost, and providing 12x more spins. It is tempting to think 
that the area comparison is unfair because ASA offers a large number of 
nominal spins. We will show next that the nominal number of spins is an 
extremely poor, if not useless, metric.

Finally, the last column shows the result of a conventional computer running
simulated annealing for reference. We can clearly see the potential benefit of 
a high-performance Ising machine: we can obtain a speedup of about 7-8 orders 
of magnitude.

\subsubsection{\bf Recap} We see that: \ding{172} Ising machines can provide
extraordinary speedups in certain types of problems; \ding{173} but for the 
scale of problems we are discussing, room-sized machines are not advantageous; 
\ding{174} ASA, OIM, and BRIM are three possible candidates for chip-scale 
applications with perhaps different strengths. We look at these models in more 
detail next.

\subsection{Detailed analyses}
\label{ssec:detail}

\subsubsection{The impact of topology}
We hope by now it is starting to be clear that BRIM is a compelling design.
Next, we want to get into details to understand some of the intrinsic
strengths that made it a compelling design. We first discuss an important
issue in Ising machine architecture. As already mentioned, when an Ising
machine uses a more local coupling network, it has a more limited ability to
map problems. Many problems would need far more physical nodes than logical
nodes in the problem. To show this, we create a number of small graphs with
edge density of 6\%.

\begin{figure}[htp]\centering
    \includegraphics[width=0.55\textwidth]{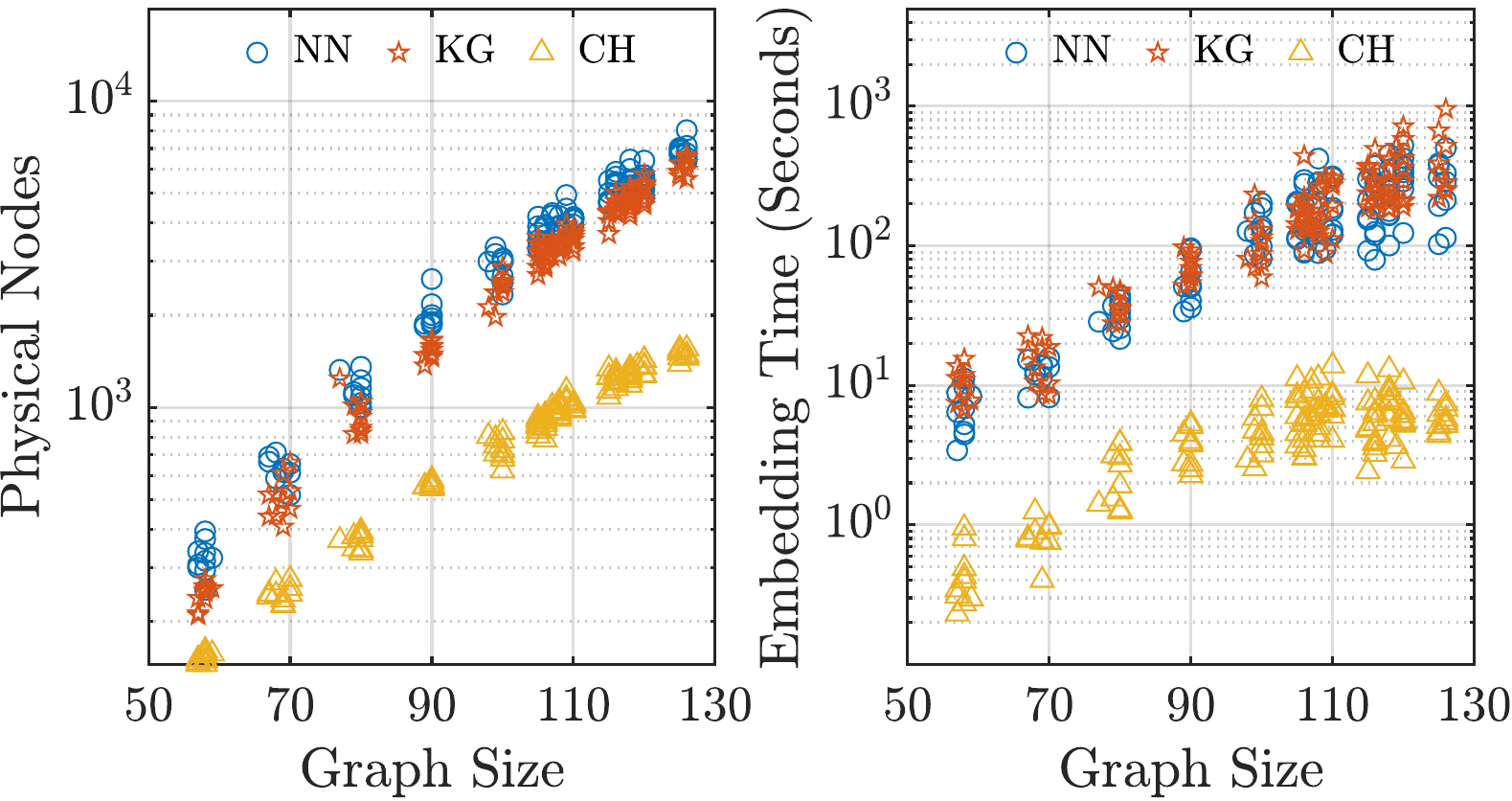}
    \caption{Minor-embedding~\cite{cai2014practical} mapping for diverse 
    graphs. Y-axis is log scale. \textbf{Left:} Total physical nodes 
    required for diverse logical node sizes. \textbf{Right:} Range of 
    embedding time. \emph{NN} is 20,000-spin Nearest-Neighbor 
    \cite{yamaoka_isscc2015}, \emph{KG} is 30,927-spin Kings-Graph
    \cite{takemoto_isscc2019}, \emph{CH} is 2048-spin Chimera.}
    \label{fig:embedding}
\end{figure} 

As shown in \fig~\ref{fig:embedding}, even with these sparse graphs, it 
takes about 1600 nodes (in D-Wave's Chimera configuration) to map a problem 
of size 126. Moreover, many nominally unused spins cannot be used due to their
location. In fact, no size 127 problems we tried can be mapped to the 2048
nominal nodes. To map these graphs, it takes up to 10 seconds to carry out the
minor embedding algorithm for Chimera. For ASA's nearest-neighbor or King's
graph topologies, the problem is more pronounced. To map the same 126-node
graph, it takes up to 8031/6649 physical nodes respectively and 
500s/955s of execution time respectively.
Note that the preprocessing time is orders of magnitude longer than the
annealing process itself and more than needed to get a better solution using
conventional simulated annealing. Finally, auxiliary
spins degrade the machine's ability to find good solutions. As we saw in
Table~\ref{tab:comp}, D-Wave and ASA have trouble reaching the ground state
for small graphs. All in all, at the moment, we see no evidence that using
local connections (at least these specific ones) is helpful in any practical
way.

\donotshow{
\todo{Something about broken nodes}

\begin{figure}[htp]
    \centerline{\includegraphics[width=0.48\textwidth]{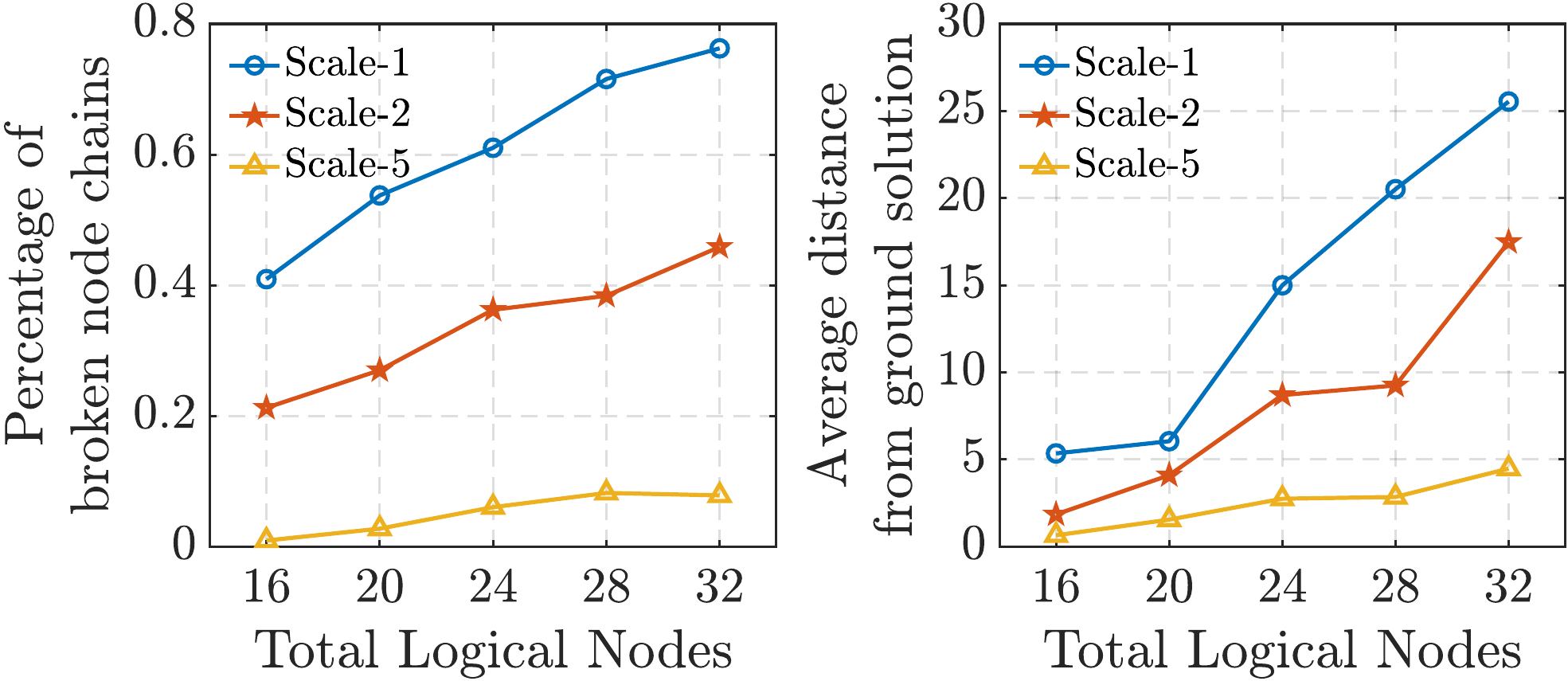}}
    \caption{Effect of broken chain due to chimera 2048 (16x16x4x2) topology 
    choice.}
    \label{fig:broken_chains}
\end{figure} 
}

\subsubsection{In-depth comparison} 
\begin{figure}[htbp]
    \includegraphics[width=0.55\textwidth]{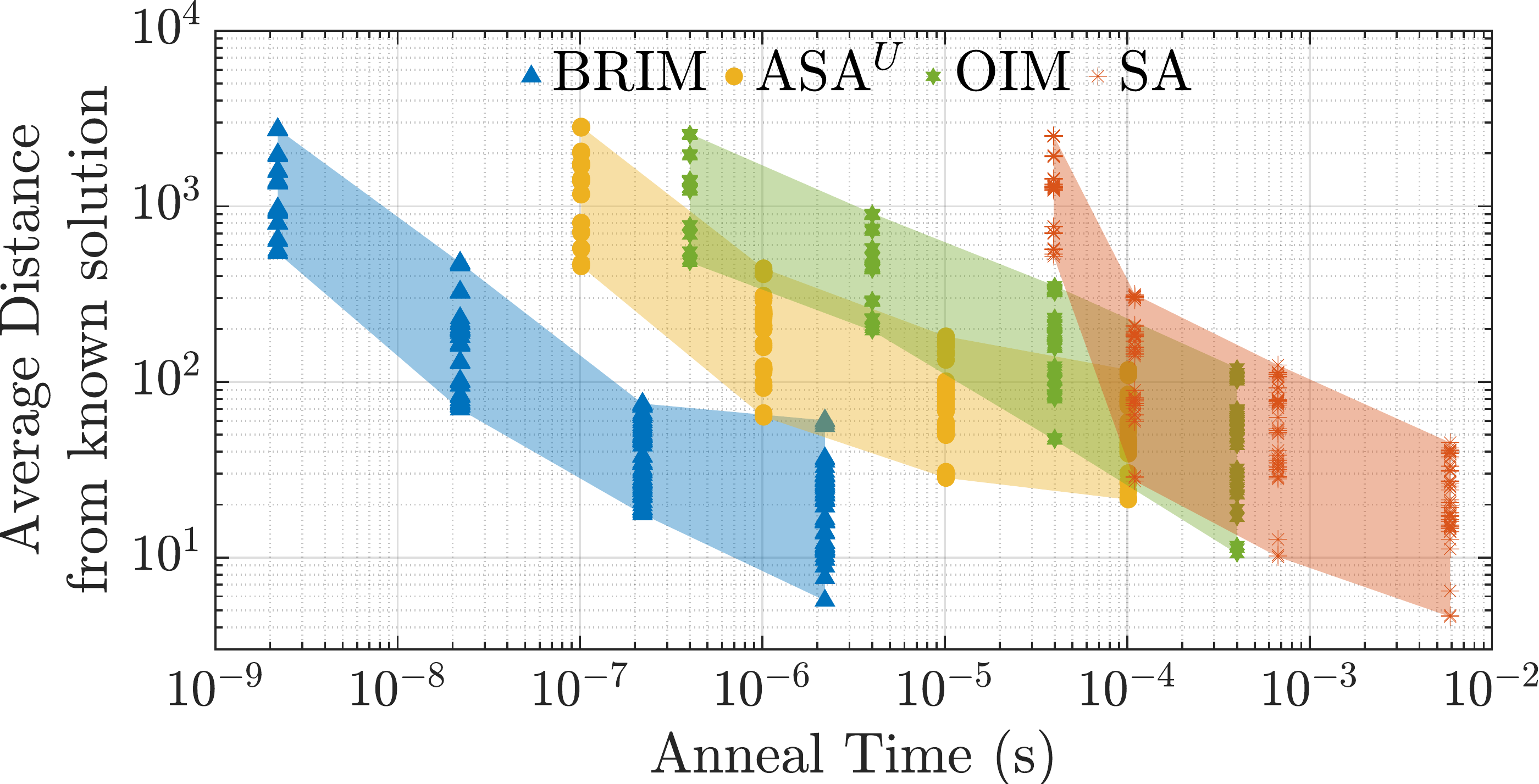}
    \caption{Solution quality measured as distance (lower is better) 
    against total annealing time. ASA$^U$ means upper-bound version 
    of the ASA hardware where we assume all-to-all connection. Each 
    marker represents the average of a single graph. For a given 
    vertical time, we show the 51 graphs.}
    \label{fig:anneal_time}
\end{figure}

Next, we study the annealing process for BRIM, ASA, and OIM more in-depth.
In \fig~\ref{fig:anneal_time}, we show the solution quality (measured as 
distance from best known solution) as different machines are given different 
annealing time.  We add simulated annealing (SA) running on a workstation 
to this mix.  Each marker's vertical coordinate represents the difference 
between the best solution and the final cut value (lower is better), while 
its horizontal coordinate shows the annealing time. The range of annealing 
times are chosen such that the quality of the solution fits into roughly 
the same band.\footnote{ Note that each additional dot to the right takes 
10x longer to obtain.} 

An important point needs to be made about ASA.~\donotshow{The 30,000 
nominal spins can \emph{not} accommodate G22's 2000 logical nodes. We 
thus create two variants instead, both better than the actual design.} 
We create an idealized model where it supports all-to-all connection 
and can update all nodes \emph{within a cohort} sequentially in one 
annealing step. Note that this is intended as a very loose upper-bound of 
performance for the same \emph{kind} of machine, and labeled as $ASA^U$ 
in \fig~\ref{fig:anneal_time}.
\donotshow{Second, 
we take the reported improvement factor ($2.6\times10^4$ in reference 
\cite{takemoto_isscc2019}) over simulated annealing and draw a dotted 
line ($ASA^R$) to indicate what a more realistic performance roughly is. 
Even this is some idealization. The reported performance improvement is 
over a much weaker graph that can be mapped onto the actual ASA. For G22, 
a much bigger system is needed and it will most definitely converge more 
slowly.}

We can see the general trend that a longer annealing time provides a 
better answer for all machines. It is also clear that BRIM is orders 
of magnitude faster than other systems. However, rather than thinking 
about these curves as fixed and precise, it is helpful to think about 
them as a general shape that can move horizontally. Their position 
only represents the effect of current design parameters. And it is 
instructive to understand \emph{what it takes} to move them to the 
left by, say, one order of magnitude. For ASA and SA, this is equivalent 
to making the machine 10x faster. We can imagine the challenge of this 
given the difficulty of scaling computation speed, not to mention memory 
access time. For the analog implementations, the question is more subtle. 
By reducing capacitance and/or inductance, the curve can shift to the left. 
The real question becomes the impact of noise and parasitics both to speed 
and solution quality. Such investigation is part of our future work. 

\subsubsection{Effects of annealing control}
\begin{figure}[tbhp]\centering
    \includegraphics[width=0.8\textwidth]{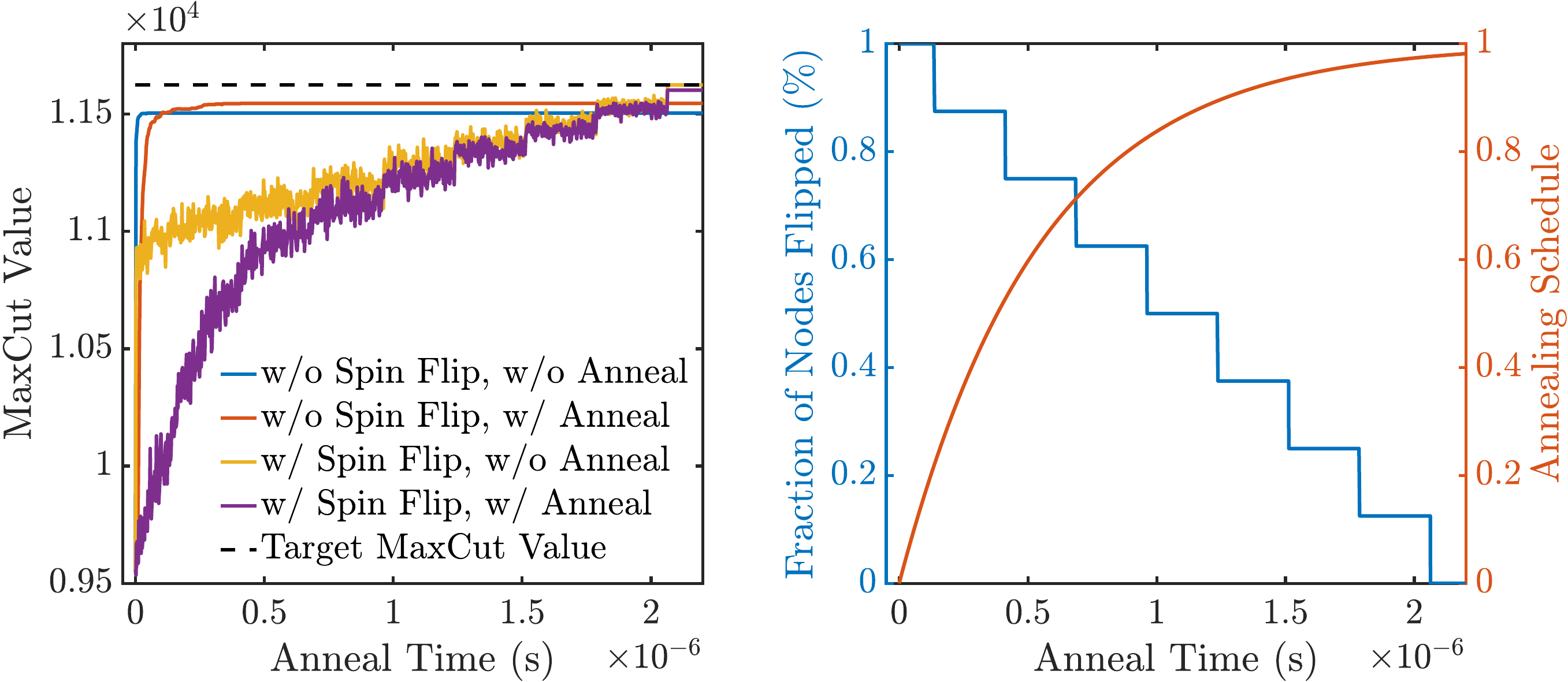}
    \caption{Diverse effects of annealing control. \textbf{Left}: Baseline 
    BRIM improvements with spin flip and/or coupling strength annealing. 
    All trajectories start at the same initial condition. \textbf{Right}: 
    Application of random spin flips (left axis) and coupling strength annealing schedule (right axis).}
    \label{fig:shock}
\end{figure}

As already mentioned earlier, this paper highlights a different approach
to Ising machine. The specific example used so far is but one design point
in the space and much of the space remains to be explored in more detail. 
For example, annealing control plays a crucial role in obtaining good 
solutions. \fig~\ref{fig:shock} compares the system with a few different 
configurations aiming at achieving better results. For example, we can
exponentially adjust the coupling strength over time. This allows the system 
to navigate the landscape at a coarser granularity early on. Additionally, 
we can periodically and randomly reverse the polarity of some nodes (to apply
random spin flips) to escape a local minimum. On the left, we show the effect 
of applying spin flips and coupling strength annealing. We find that applying
random spin flips are indeed very effective as expected.\footnote{Instead of 
adding completely random spin flips, we also studied another approach in 
simulation. In this approach, the energy values of the current state and
destination state are calculated and whether the destination state is accepted
depends on the energy values as in a Metropolis–Hastings algorithm. We note 
that this approach does offer better results though it is challenging to 
implement in hardware.}

\subsubsection{Effects of bit precision in coupling control}

Since BRIM uses a number of analog circuits, it is natural to be concerned about 
the impact
of noise and other non-idealities on its operation. Building and testing actual
hardware will be the ultimate experiment to answer those concerns. But a number
of simulation-based experiments can still shed some light. We first look at
whether we can provide enough precision in programming the coupling weights.

\begin{figure}[htp]\centering
\includegraphics[width=0.7\textwidth]{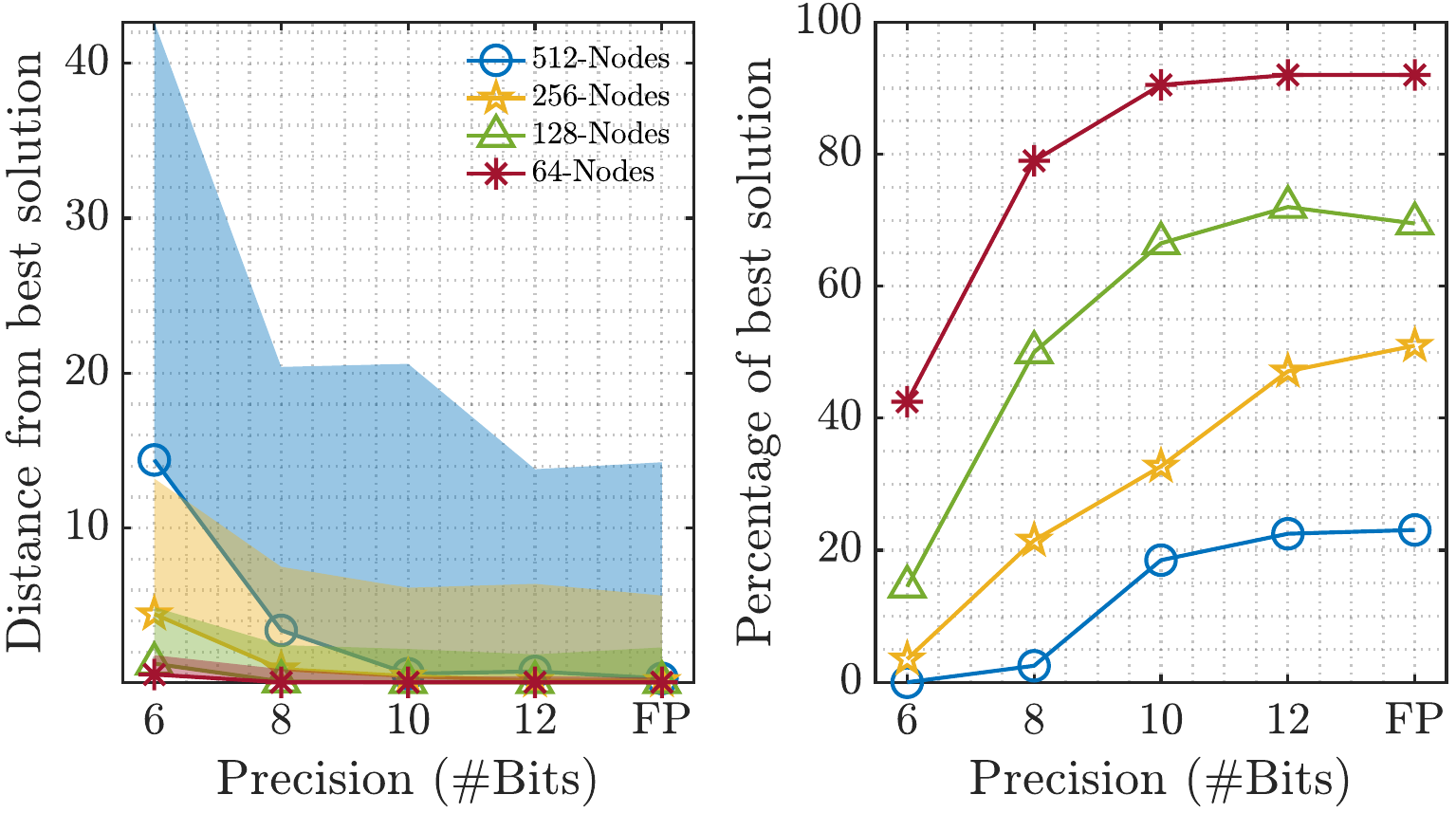}
\caption{BRIM solution quality at diverse bit precision. Graphs have all-to-all 
connection with random weights uniformly distributed in [-1, 1]. Solid lines are
averages, and shaded region is range.}
\label{fig:num_bits}
\end{figure}

To answer the question, we compare BRIM with different bit precisions in the weight
matrix. The Gset benchmarks actually have weights of $\pm1$. So for this experiment,
We introduce a hypothetical problem 
with real coupling weights uniformly distributed in [-1, 1]. We then set the 
programming system including DAC to support 6 to 12 bits of precision and compare
the result to the reference of supporting single-precision floating point weights.
\fig~\ref{fig:num_bits} shows the range of the solution on the left and the
average probability of achieving ground state solution on the right.

We can see that dropping weight 
precision from single-precision FP to 12 bits has virtually no impact in any
respect. Further reducing the precision will bring some noticeable impact, with
6 bits being on the boundary of acceptability. Keep in mind, such uniformly
distributed weights represent perhaps the extreme of demand on weight precision.

\subsubsection{Effects of variation and noise}
To evaluate the effect of device variations we introduced mismatch to coupling 
resistors ($\mu=0,\sigma=3\%$), buffer gain ($\mu=1,\sigma=3\%$), and buffer 
offset ($\mu=0\,mV,\sigma=10\,mv$) drawn from normal distribution. Additionally,
we also introduced electronic noise ($\mu=0\,mV,\sigma=0.35\,mV$) to the output
of every node over the entire annealing process. 
The magnitude and variation of the noise are
obtained from circuit simulation by integrating the output noise at 
$out+$ and $out-$ of Fig~\ref{fig:coupling_unit} from within the bandwidth of 
the voltage buffer (1Hz to 100MHz).
Overall, the impact of noise and variation is rather tolerable. 
This can be seen by the average distance to the 
best solution. Without variation and noise, 
the distance is 3.8. Accounting for device variations increases the distance 
to 4.2. Adding noise, with or without variations, increases the distance to 5.

\donotshow{
\subsubsection{Scaling and Effects of Delay}
BRIM area scales roughly at $N^2$  -- the problem description itself (coupling 
matrix) scales at $N^2$ and state space scales at $e^N$. For dense problems, 
no Ising machines can scale better than $N^2$. Considering effective (not 
nominal) spins, local-connection machines (D-Wave, ASA) scale worse. Propagation 
delay is more complicated. But at chip-level the propagation delay (10s of ps) 
is negligible compared to nodal time constants (at least ns). We simulate the
effect of delay using MATLAB's Delay Differential Equations (DDE) with up to 
$1\,ps$ delay. We observed the average solution quality increase from baseline 
$3.8$ to $5$.
}

\subsection{Discrete-component prototype}
\label{sec:BRIM prototype}

We now showcase a limited prototype of BRIM using discrete electronics
with operational amplifiers and passive components such as capacitors as well
as resistors. The prototype is meant to demonstrate the key operation principle
and therefore does not have some features such as programming interface and
advanced annealing control to escape local minima.

\begin{figure}[htp]
    \centering
    \includegraphics[width=0.6\textwidth]{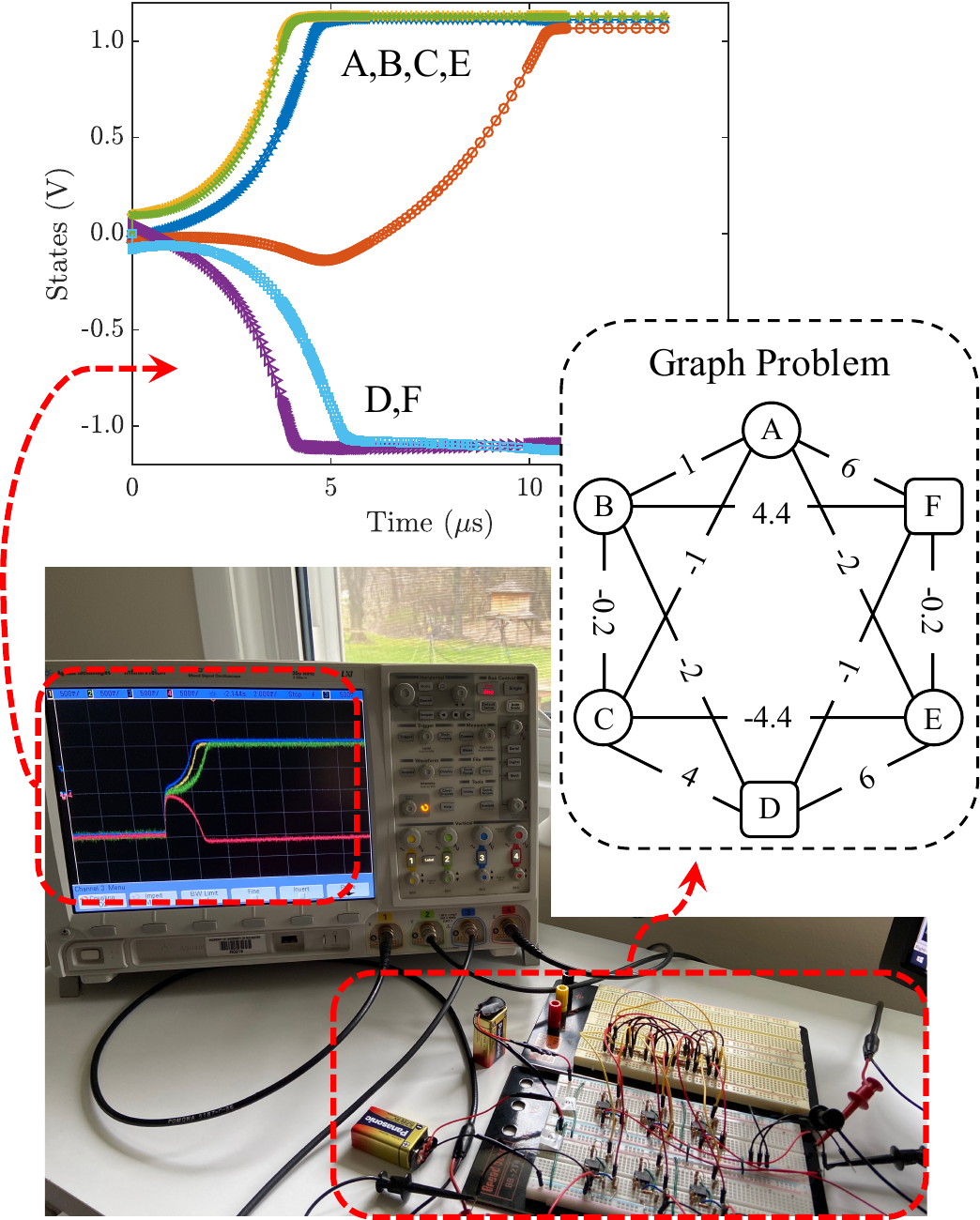}
    \caption{Photo of the discrete-component prototype (with bi-directional
    coupling) connected to a \emph{4-channel scope} after powering up (bottom). 
    Note that scope captures 4 of the 6 bifurcation nodes ($A$ to $F$). The 
    graph representation of the problem being solved and a Spice-simulated 
    waveform are also shown (top). In this particular simulation, nodes $A$, 
    $B$, $C$, $E$ converge to $+1.15V$ while the other two nodes converge 
    to $-1.15V$ representing an opposite spin.}
    \label{fig:discrete}
\end{figure}

The six nodes are implemented each with a capacitor and a ZIV diode made
out of an op-amp. These are seen in the bottom right of \fig~\ref{fig:discrete}. 
The coupling can be seen above the nodes in the figure. In this prototype, 
the nodes are directly connected with resistors without any buffer. Thus the 
coupling is bidirectional.

There is no state initialization in this prototype and each node starts at
a voltage of 0. Noise intrinsic to the setup helps each node break out of
the saddle point ($v_i = 0$) and bifurcate to one of the stable states of
$\pm 1.15V$ determined by the specific set up of the ZIV diodes. Each run
from power-up thus gives a different waveform on the scope and a potentially
different solution. A waveform from simulation as well as the waveform on the
scope can be see in \fig~\ref{fig:discrete}. In this particular setup, the
energy landscape is relatively simple and the prototype always reaches 
the ground state.

\section{Conclusions}\label{sec:conclusion}

Ising machines can be programmed to map an abstract problem and let physics
naturally guide the dynamical system towards some kind of optimal state.
This state can then be translated
back to be a heuristic solution to a combinatorial optimization problem. The
use of nature suggests the possibility of significant speed and/or efficiency.
Consequently, exploration of Ising machines has been gaining attentions.
Quantum annealers and optical coherent Ising machines are prominent examples
of such machines which drew particular interests from the physics community.
However, these Ising machines are generally bulky and energy intensive.
Continued exploration is certainly a worthwhile endeavor for scientific
purposes. But for now, integrated circuit designs allow more immediate
applications. We propose one such design we call BRIM that uses bistable
nodes resistively coupled with programmable and variable strengths. The design
is fully CMOS compatible for chip-scale implementations. Through our experimental
analysis, we show that the machine is significantly smaller, faster, and more
energy-efficient relative to some other designs.
Compared to the room-sized Ising machines, it is about 4 
orders of magnitude faster and consuming 7-8 orders of magnitude less energy.
Compared to a recent chip-scale simulated annealing accelerator, it is also
orders of magnitude better in speed, area, and power. While there is
some degree of uncertainty with these statistics, we also envision continued
improvement and optimizations. On the other hand, there are also many
challenges facing machines like BRIM. Scalable architecture, area-efficient
connection networks, and circuit scaling in the presence of Process-Voltage-Temperature 
(PVT) variations are but a few examples of the challenges.

\bibliographystyle{ACM-Reference-Format}
\bibliography{main}
\end{document}